\documentclass[useAMS,usenatbib]{mn2e}
\usepackage{graphicx}
\usepackage{times}
\usepackage{amssymb}
\title[Cosmological Constraint from Large Scale Weak Lensing]{Cosmological Constraints from the Large Scale Weak Lensing of SDSS MaxBCG Clusters}
\author[Zu et al.]{ 
\parbox{\textwidth}{
Ying Zu$^1$\thanks{E-mail: yingzu@astronomy.ohio-state.edu},
David H. Weinberg$^{1}$,
Eduardo Rozo$^{2,3}$, 
Erin S. Sheldon$^{4}$, 
Jeremy L. Tinker$^{5}$, 
Matthew R. Becker$^{3,6}$
}
\vspace*{4pt} \\
${1}$~Department of Astronomy and CCAPP, The Ohio State University, 140 W. 18th Avenue,
Columbus, OH 43210, USA\\
${2}$~Einstein Fellow, Department of Astronomy \& Astrophysics, The University of Chicago, Chicago, IL 60637, USA\\
${3}$~Kavli Institute for Cosmological Physics, 5640 South Ellis Avenue, The University of Chicago, Chicago, IL 60637, USA\\
${4}$~Brookhaven National Laboratory, Bldg 510, Upton, New York 11973, USA\\
${5}$~Center for Cosmology and Particle Physics, Department of Physics, New York University, USA\\
${6}$~Department of Physics, 5720 S. Ellis Avenue, The University of Chicago, Chicago, IL 60637, USA\\
}
\def\M{M} % mass, turns out not necessary
\def\mrm{\mathrm}
\def\rich{N_{200}}
\def\om{\Omega_m}
\def\s8{\sigma_8}
\def\sc{\sigma_{\ln {\rich|\M}}}
\def\be{\beta}
\def\la{\ln \bar{N_{1}}}
\def\lb{\ln \bar{N_{2}}}
\def\lc{\ln \bar{N_{3}}}
\def\ma{\M_{1}}
\def\mb{\M_{2}}
\def\mc{\M_{3}}
\def\dns{\Delta{n_{s}}}
\def\ns{{n_{s}}}
\def\scm{\sigma_{\ln\M|\rich}}
\def\lcdm{\Lambda\mbox{CDM}}
\def\ob{\Omega_{b}}
\def\on{\Omega_{\nu}}
\def\mpc{\mathrm{Mpc}}
\def\hmpc{h^{-1}\mathrm{Mpc}}
\def\hkpc{h^{-1}\mathrm{kpc}}

\def\pk{P(k)}
\def\plin{P_\mathrm{lin}(k)}
\def\pnl{P_\mathrm{nl}(k)}
\def\preid{P_\mathrm{Reid10}(k)}

\def\pks{P(k)\;\mbox{shape}}
\def\msun{\M_\odot}
\def\hmsun{h^{-1}\M_\odot}

\def\ds{\Delta\Sigma}
\def\ltail{\mathcal{L}_\mathrm{tail}}
\def\zph{z_\mathrm{photo}}
\def\frbin{\psi(\rich)}

\def\frbinl{\psi_l(\rich)}
\def\fzphbin{\phi(\zph)}
\def\frbinm{\langle\psi |\M\rangle}

\def\frbinml{\langle\psi_l |\M\rangle}
\def\fzphbinz{\langle\phi |z\rangle}
\def\wgt{\omega(\M, z)}

\def\wgtj{\omega_j(\M, z)}

\def\wgtl{\omega_l(\M, z)}

\def\xihm{\xi_\mathrm{hm}}
\def\xicm{\xi_\mathrm{cm}}
\def\ximm{\xi_\mathrm{mm}}
\def\xilin{\xi_\mathrm{lin}}
\def\xinl{\xi_\mathrm{nl}}
\def\cov{C}
\def\Rmin{R_\mrm{min}}
\def\covnn{\cov_\mrm{NN}}
\def\covnnii{\cov_{\mrm{N}^i\mrm{N}^{i'}}}
\def\covnniidiag{\cov_{\mrm{N}^i\mrm{N}^{i}}}
\def\covns{\cov_\mrm{\mrm{N}\Sigma}}
\def\covss{\cov_\mrm{\Sigma\Sigma}}

\def\avg#1{\left\langle #1 \right\rangle}
\begin{document} 
%\fontsize{12pt}{14pt}\selectfont
\date{\today} \maketitle
%%%%%%%%%%%%%%%%%%%%%%%%%%%%%%%%%%%%%%%%%%%%%%%%%%%%%%%
%
%%%%%%%%%%%%%%%%%%%%%%%%%%%%%%%%%%%%%%%%%%%%%%%%%%%%%%%
%%  Abstract and Keywords
%%%%%%%%%%%%%%%%%%%%%%%%%%%%%%%%%%%%%%%%%%%%%%%%%%%%%%%
\begin{abstract}
We derive constraints on the matter density $\om$ and the amplitude of matter
clustering $\s8$ from measurements of large scale weak lensing (projected
separation $R=5-30\hmpc$) by clusters in the Sloan Digital Sky Survey MaxBCG
catalog. The weak lensing signal is proportional to the product of $\om$ and
the cluster--mass correlation function $\xicm$. With the relation between
optical richness and cluster mass constrained by the observed cluster number
counts, the predicted lensing signal increases with increasing $\om$ or $\s8$,
with mild additional dependence on the assumed scatter between richness and
mass. The dependence of the signal on scale and richness partly breaks the
degeneracies among these parameters.  We incorporate external priors on the
richness--mass scatter from comparisons to X-ray data and on the shape of the
matter power spectrum from galaxy clustering, and we test our adopted model for
$\xicm$ against N-body simulations. Using a Bayesian approach with minimal
restrictive priors, we find $\s8(\om/0.325)^{0.501}=0.828\pm0.049$, with
marginalized constraints of $\om=0.325_{-0.067}^{+0.086}$ and
$\s8=0.828_{-0.097}^{+0.111}$, consistent with constraints from other MaxBCG
studies that use weak lensing measurements on small scales~($R \leq 2\hmpc$).
The $(\om,\s8)$ constraint is consistent with and orthogonal to the one
inferred from WMAP CMB data, reflecting agreement with the structure growth
predicted by General Relativity for a $\Lambda$CDM cosmological model. A joint
constraint assuming $\Lambda$CDM yields $\om=0.298_{-0.020}^{+0.019}$ and
$\s8=0.831_{-0.020}^{+0.020}$. For these parameters and our best-fit scatter we
obtain a tightly constrained mean richness-mass relation of MaxBCG clusters,
$N_{200}=25.4(M/3.61 \times 10^{14} \hmsun)^{0.74}$, with a normalization
uncertainty of $1.5\%$  Our cosmological parameter errors are dominated by the
statistical uncertainties of the large scale weak lensing measurements, which
should shrink sharply with current and future imaging surveys.
\end{abstract}
%%%%%%%%%%%%%%%%%%%%%%%%%%%%%%%%%%%%%%%%%%%%%%%%%%%%%%%
%%%%%%%%%%%%%%%%%%%%%%%%%%%%%%%%%%%%%%%%%%%%%%%%%%%%%%%
%%%%%%%%%%%%%%%%%%%%%%%%%%%%%%%%%%%%%%%%%%%%%%%%%%%%%%%
\section{Introduction}
\label{sec:intro}

The most fundamental question about the origin of cosmic acceleration is
whether it arises from a new energy component or from a modification of General
Relativity~(GR) on cosmological scales. A general strategy to address this
question is to compare the growth of cosmic structure --- as measured, e.g., by
cosmic shear, redshift--space distortions of galaxy clustering, or the
abundance of galaxy clusters as a function of mass --- to the predictions of a
GR$+$dark energy model constrained by geometrical probes such as Type Ia
supernovae and baryon acoustic oscillations~(BAO).  In particular, one can
compare measurements of the matter density $\om$ and the present--day amplitude
of matter clustering, characterized by $\s8$, the rms matter fluctuation in
$8\hmpc$ spheres, to the values expected from extrapolating cosmic microwave
background~(CMB) anisotropies forward from recombination to $z=0$.\footnote{We
define $h\equiv \mrm{H}_0/(100\,\mrm{km}\,\mrm{s}^{-1}\mrm{Mpc}^{-1})$ where
$\mrm{H}_0$ is the Hubble parameter at $z=0$.} Cosmological studies with
clusters traditionally use mass proxies derived from X-ray or
Sunyaev--Zel'dovich~\citep[SZ;][]{sunyaev1972} measurements to constrain the
cluster mass function $dn/d\M$~(see \citeauthor{allen2011} 2011 for a review).
In an alternative approach, \citeauthor{sheldon2009}~(2009; hereafter S09) used stacked
weak lensing~(WL) to measure the average mass profiles around clusters in the
MaxBCG catalog~\citep{koester2007} derived from the Sloan Digital Sky
Survey~(SDSS; \citeauthor{york2000} 2000), detecting correlated mass from
scales of $0.1\hmpc$ to $30\hmpc$. \citeauthor{rozo2010}~(2010; hereafter R10) used the
S09 measurements to constrain the mean relation between optical richness and
virial mass for MaxBCG clusters, and they combined this relation with the
abundance of clusters as a function of richness to constrain $\om$ and $\s8$.
(For a general review of this approach in the context of cluster cosmology, see
\S6 of \citeauthor{weinberg2012} 2012.) In this paper we again target $\om$ and
$\s8$ with MaxBCG clusters, but we use the {\it large scale} S09 measurements,
from projected separations of 5--30$\hmpc$.

Roughly speaking, stacked weak lensing measures the product of the matter
density  $\om$ and the cluster--mass cross--correction function $\xicm(r)$.
More precisely, given knowledge of the distances to lensing clusters and
background sources, the mean tangential shear profile of clusters measures the
excess surface density profile $\ds(R)$, which is related to the 3--d
$\xicm(r)$ via
\begin{eqnarray}
\ds(R)  & = &
\om\rho_{\mrm{c}}\frac{2}{R^2}\int_0^{R}\int_{-\infty}^{+\infty}r_p\xicm\left(\sqrt{r_p^2+r_z^2}\right)dr_zdr_p
\nonumber
\\ & - &
\om\rho_{\mrm{c}}\int_{-\infty}^{+\infty}\xicm\left(\sqrt{R^2+r_z^2}\right)dr_z
\end{eqnarray}
(see~\S~\ref{sec:data} for further details).  We can understand how large scale
$\ds(R)$ measurements constrain $\om$ and $\s8$ by considering the simple case
in which optical richness is perfectly  correlated with cluster mass, so that a
sample of clusters above a richness threshold corresponds to a sample above a
mass threshold that has the same comoving space density $\bar{n}$~(where, for
simplicity, we consider a sample at fixed redshift). For a given cosmological
model, one can predict the matter correlation function $\ximm(r)$ and the bias
factor $b_c(\bar{n})$ of halos with space density $\bar{n}$, and thus the
cluster--mass correlation function, which is $\xicm(r) = b_c(\bar{n})\ximm(r)$
on scales in the linear regime. Raising $\om$ with all other quantities held
fixed raises the predicted $\ds(R)$ proportionally. Raising $\s8$ increases
$\ximm\propto\s8^2$ and thus increases $\xicm(r)$, but there is a partly
compensating decline in $b_c(\bar{n})$. In the limit of very rare, very highly
biased peaks, $b_c(\bar{n})\propto\s8^{-1}$, yielding $\ds(R)\propto\om\s8$,
but for the space densities of typical cluster samples $b_c(\bar{n})$ drops
more slowly than $\s8^{-1}$. Thus, the combination of cluster abundance
measurements, which determine $\bar{n}$, and large scale weak lensing
measurements, which determine $\ds(R)$, constrains a parameter combination
$\s8\om^\gamma$ with $\gamma<1$. In practice, we will use bins of cluster
richness instead of a single sample above a threshold, and the
$\s8$--dependence  of $\xicm(r)$ is different in the linear and mildly
non--linear regimes, so there is some leverage to break degeneracy between
$\om$ and $\s8$.

The simplifications of this description point up several complications that must
be addressed in our analysis. First, the measurements of $\ds(R)$ have
systematic uncertainties related to the photometric redshifts of the sources and
shear calibration. Second, optical richness is a mass indicator with substantial
scatter, which makes the bias of clusters in richness bins different from that
of mass bins with the same space density. The two principal ``nuisance
parameters'' in our statistical analysis are $\be$, an overall scaling of the
$\ds(R)$ measurements to allow for systematic uncertainty, and $\sc$, the
logarithmic scatter in richness at fixed mass. We discuss these nuisance
parameters and the priors we adopt on them in~\S~\ref{sec:ana}. We also adopt a
prior on the shape of the matter power spectrum, so that a value of $\s8$
specifies the full shape of $\ximm(r)$. The inference of comoving space
densities itself depends on $\om$, which affects the volume element
transformation between comoving distances and observable angles and redshifts.
Incompleteness and contamination of the cluster sample can also affect the
inferred space densities and/or bias the estimate of $\ds(R)$, so they must also
be accounted in the analysis. Despite these complications, we find that our
constraints are limited by the statistical errors of the weak lensing
measurements rather than systematic uncertainties.

A complete cosmological analysis of cluster weak lensing would employ $\ds(R)$
measurements over the full range of observed scales. Here we restrict our
analysis to $R\geqslant5\hmpc$, in part to avoid the regime where theoretical
predictions of $\xicm(r)$ are uncertain, and in part to keep our results
complementary to those of R10, who use the small scale ($R\lesssim2\hmpc$) S09
measurements to calibrate their determination of the cluster mass function. One
important systematic for interpretation of the small scale measurements is the
impact of cluster mis--centering, which must be estimated from simulations of
the cluster population and cluster finding
technique~\citep[e.g.][]{johnston2007, george2012}. One advantage of the
approach in this paper is that mis--centering has negligible impact at the
large scales that we employ.

\begin{figure*}
\centering
\resizebox{1.0\textwidth}{!}
{\includegraphics{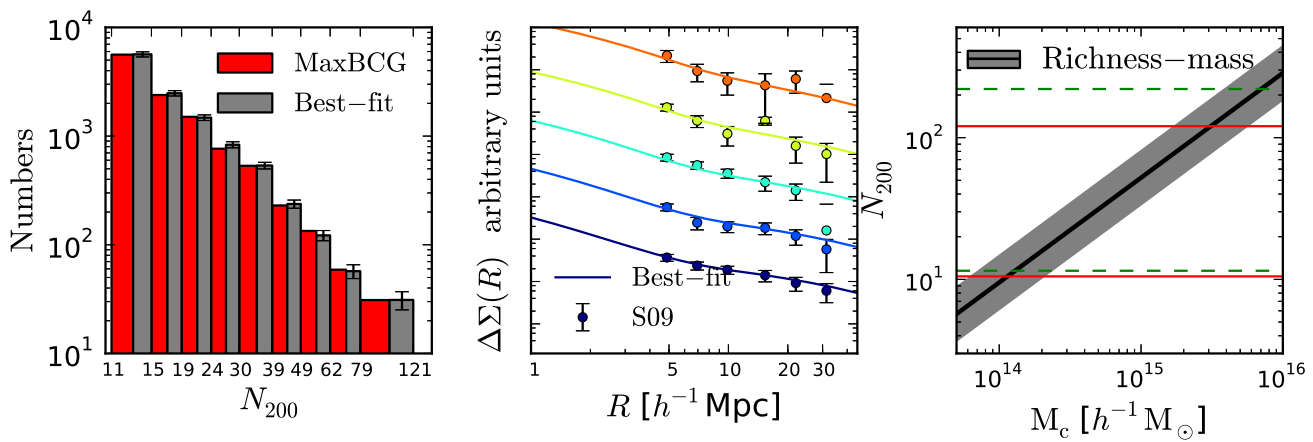}}
\caption{ Comparison between observables used in the analysis and the best--fit model
predictions. Left panel: Cluster number counts from the MaxBCG sample~(dark/red
histograms) and the best--fit model prediction~(light histograms with
errorbars). Middle panel: Stacked surface density contrast profiles of five
richness bins measured by S09~(solid circles with errorbars) and predicted by
the best--fit model~(solid curves), multiplied by a different constant for each
bin to avoid crowding. Errorbars in the left and middle panels correspond to the
square root of the diagonal terms in the covariance matrix.  Right panel: Mean
richness--mass relation~(thick solid line) and intrinsic scatter~(gray
shaded stripe) of the best--fit model.  Solid and dashed horizontal lines bracket
the richness ranges associated with the cluster samples we used for the number
counts and $\ds(R)$, respectively.  }
\label{fig:model_demo}
\end{figure*}

In the following section we briefly review our input data, the MaxBCG cluster
catalog and the S09 weak lensing measurements. Section~\ref{sec:ana} presents
our analysis method in detail, including the model parameters and priors and the
procedure for computing the likelihood of the data given these parameters.
Section~\ref{sec:simumock} tests our analytic models for $\ds(R)$~(a modified
version of that proposed by~\citeauthor{hayashi2008} 2008, hereafter HW08)
against numerical simulations, and it uses simple mock data sets to test other
aspects of our analysis procedures. Section~\ref{sec:results} presents our
cosmological constraints and compares them to those from other cluster analyses
and from CMB data. We address systematic uncertainties
in~\S\ref{sec:systematics}. We close, in~\S\ref{sec:dis}, with a summary of our
findings and a discussion of future prospects. The reader in a hurry can get an
overview of the paper from Fig.~\ref{fig:model_demo}, which compares our
best--fit model to our input data, Fig.~\ref{fig:pedagogical}, which shows how
the $\ds(R)$ prediction depends on model parameters, and Fig.~\ref{fig:wmap},
which presents our derived constraints on $\om$ and $\s8$.

\section{Data}
\label{sec:data}

\subsection{Cluster Catalog and Number Counts}
\label{sec:abundance_data}

The MaxBCG cluster catalog~\citep{koester2007} consists of $13,823$ clusters
identified from the imaging data of the SDSS Data Release
4~\citep[DR4;][]{adelman-mccarthy2006}. The clusters are selected as spatial
overdensities of red galaxies, which form a tight E/S0 ridgeline in the
color--magnitude diagram. Each cluster is assigned a richness measure $\rich$,
defined as the number of red--sequence galaxies with $L>0.4L_*$ in the {\it
i}--band within a scaled radius $R_{200}$ such that the galaxy density interior
to that radius is $200$ times the mean galaxy density. The tight relation
between the ridgeline color and redshift also allows an accurate photometric
redshift estimate for each cluster~($\Delta z \simeq 0.01$). Tests against mock
catalogs suggest MaxBCG is $\sim 90\%$ complete and pure for clusters with
masses $\geq 10^{14}\hmsun$, and  $95\%$ for higher mass
clusters~\citep{rozo2007-1}. The
MaxBCG catalog is nearly volume--limited in the redshift range between $0.1$ and
$0.3$ over $7398$ deg$^2$. The large volume and dynamic range in mass make it
well suited to our cosmological analysis.

For better control on purity and completeness, we only keep $10,815$ MaxBCG
clusters~($78\%$ of total) with $\rich > 11$ for the abundance measurement.
Since there are only $5$ clusters with $\rich > 120$ but they span a large richness
range to $\rich^\mathrm{max} = 188$, we need to model the number counts for
clusters with $\rich$ below and above $120$ separately. We will describe the
difference of their treatments in our likelihood model in~\S~\ref{sec:like}.  

For the clusters with $\rich \in \{11\cdots120\}$, Table~\ref{tab:abundancebins}
gives our richness binning, and the red histograms in the left panel of
Fig.~\ref{fig:model_demo}~(discussed further below) show the measured number
counts in those bins.  When predicting these numbers for a set of model
parameters, we integrate over redshift and account for scatter between
photometric and true redshift, with a survey area of $7398$ deg$^2$. To a good
approximation, the predicted counts are what one would obtain using the halo
mass function at $z=0.23$ and the comoving volume from $z=0.1$ to $z=0.3$.

\subsection{Large Scale Cluster Weak Lensing Measurements}
\label{sec:wl_data}

We take our large scale weak lensing measurements from S09, who measured the
mean tangential shear profiles $\gamma_T(R)$ of source galaxies around lens
clusters in bins of richness. The area of the imaging data used for this
analysis is somewhat smaller than the $7398$ deg$^2$ used for the number counts.
The mean tangential shear is then converted to the
mean excess surface density profile $\ds(R)$ for each richness bin,
\begin{equation}
\ds(R) \equiv  \bar{\Sigma}(<R) - \bar{\Sigma}(R) = \gamma_T(R) \times \Sigma_\mrm{crit},
\end{equation}
where $\bar{\Sigma}(R)$ is the azimuthally averaged density at projected radius
$R$ and $\bar{\Sigma}(<R)$ is the mean surface density {\it interior} to
$R$~\citep[see, e.g.][]{miralda-escude1991, sheldon2004}. The critical surface
density $\Sigma_\mrm{crit}$ above is defined to be
\begin{equation}
\Sigma_\mrm{crit} = \frac{c^2}{4\pi G}\frac{D_S}{D_{LS}D_L},
\end{equation}
where $D_S$, $D_L$, and $D_{LS}$ are the angular diameter distances from the
observer to the source, from the observer to the lens, and between the lens and
source, respectively. To calculate $\Sigma_\mrm{crit}$, S09 estimated $D_S$ and
$D_{LS}$ using the photo--{\it z}'s of source galaxies, so any uncertainties
in the photo--{\it z} estimates affect the measurements of $\ds(R)$, as we will
describe in \S~\ref{sec:par}. The values of $\Sigma_\mrm{crit}$ are computed for
a spatially flat universe with $\om=0.28$ and a cosmological constant. Over our
redshift range, the impact of varying this assumption is negligible compared to
the statistical errors, so we do not adjust $\Sigma_\mrm{crit}$ when fitting
cosmological parameters.

While the signal-to-noise ratio of $\ds(R)$ for each cluster is small, S09
stacked the signals among all clusters in each richness bin to obtain average
$\ds(R)$ profiles. The stacked signal was detected from the inner
halo~($25\hkpc$) well into the surrounding large scale structure~($30\hmpc$). As
mentioned in the introduction, the small scale measurements were used by R10 for
their constraints, and we hope to employ the large scales by interpreting
$\ds(R)$ as a measure for $\om\xicm$.

Table~\ref{tab:dsrbins} summarizes the richness binning for $\ds(R)$, and the
solid circles in the middle panel of Fig.~\ref{fig:model_demo} show the measured
$\ds(R)$ on large scales. Errorbars are derived from jackknife
re--sampling~(see~\S~\ref{sec:like} below) and are correlated between points. We
take our data values for $\ds(R)$ from table 1 of S09 with one important
correction. As first noted by~\cite{mandelbaum2008}, the weak lensing signal in
S09 appears to be diluted because of photometric redshift errors that
incorrectly locate some foreground galaxies~(which cannot be lensed by the
clusters) behind the clusters. \cite{rozo2009} estimated that the original S09
measurements of $\ds(R)$ should be multiplied by factor $1.18$, with uncertainty
of $0.04$. We adopt this factor of $1.18$ to scale the original S09 $\ds(R)$ and
its associated error matrix up as the actual data for the analysis, and we refer
to these scaled data as the ``S09 measurements'' in the rest of the paper. 

\begin{table}
\centering
\caption{Richness bins for the abundance data. The average mass and
bias of clusters in each bin are computed from the best--fit MaxBCG$+$WMAP7 joint model.}
\begin{tabular}{cccc}
\hline
Richness & No. of Clusters $N_i$ & $\avg{\M_{200m}}_i$ [$\hmsun$] & $\avg{b}_i $ \\
\hline
11-14  & 5167 & $0.997\times10^{14}$&  2.373 \\
15-18  & 2387 & $1.404\times10^{14}$&  2.696 \\
19-23  & 1504 & $1.857\times10^{14}$&  3.010 \\
24-29  & 765  & $2.417\times10^{14}$&  3.359 \\
30-38  & 533  & $3.153\times10^{14}$&  3.772 \\
39-48  & 230  & $4.108\times10^{14}$&  4.260 \\
49-61  & 134  & $5.200\times10^{14}$&  4.769 \\
62-78  & 59   & $6.556\times10^{14}$&  5.353 \\
79-120 & 31   & $8.627\times10^{14}$&  6.170 \\
\hline
\end{tabular}
\label{tab:abundancebins} 
\end{table}

\section{Analysis}
\label{sec:ana}

To obtain constraints on $\om$ and $\s8$, we
adopt a Bayesian approach with a minimal set of restrictive priors on nuisance
parameters. To facilitate the analysis, we also utilize information known
from other experiments, specifically, the shape of the linear power spectrum and
the scatter of cluster masses at given richness, as priors into our analysis.
Details on the model parameters, likelihood components, and the prior
specifications can be found below.

\subsection{Model Parameters}
\label{sec:par}

We assume a flat $\lcdm$ cosmology and infer the values of $\om$ and $\s8$,
along with other nuisance parameters. Any deviation from the flat $\lcdm+$GR
assumption would manifest itself as inconsistent constraints on $\om$ and $\s8$
compared to expectations from the CMB, because of growth that differs from
predictions of the GR$+$cosmological constant model. Since we introduce the
shape of the linear power spectrum $\plin$ inferred from galaxy redshift surveys
as a prior, we do not assume specific values for the tilt of the primordial
power spectrum $\ns$, the baryon density $\ob h^2$ and the neutrino mass $\on
h^2$, but only require them to be consistent with the input power spectrum shape
and the output constraints on $\om$ and $\s8$. Our analysis is also independent
of the assumed value of the Hubble parameter $h$, as both the power spectrum
shape and the weak lensing shear are measured from galaxies and clusters in the
local universe, so that all distances are in units of $\hmpc$. While the $\pks$
is determined very well on the scales that are relevant to galaxies and
clusters~\citep[][hereafter Reid10]{reid2010}, it could swing away on other
scales. We therefore allow rotational freedom in the $\pks$ by introducing a
modification to the overall tilt as another parameter $\dns$, representing
residual uncertainty in the $\pks$. The final linear power spectrum $\plin$ is
then $\propto\preid k^{\dns}$, normalized accordingly by the input $\s8$. We
comment more on the $\pks$ in \S~\ref{sec:testshape}.

Following R10, we assume that the mean cluster richness--mass relation is a
power-law, parameterized by two mean log-richnesses, $\la$ and $\lb$, at $\ma =
1.3\times10^{14}\hmsun$ and $\mb = 1.3\times10^{15}\hmsun$, respectively.  We
investigate the effect of allowing deviation from a power-law in
\S~\ref{sec:systematics}.  To go from the {\it expected} mean richness of a
cluster of mass $\M$ to the actual observed richness, we assume a log--normal
distribution with a constant scatter $\sc$ across all cluster masses. Note that,
unlike R10, we use mass unit $\hmsun$ rather than $\msun$ throughout the
analysis, in accordance with our choice of distance units to avoid dependence on
$h$. The right panel of Fig.~\ref{fig:model_demo} shows the richness--mass
relation predicted by our best--fit model, with $\la=2.446$, $\lb=4.148$,
and scatter $\sc=0.432$~(gray shaded band). The solid and dashed vertical lines
indicate the richness ranges of clusters used in the number count and weak
lensing constraints, respectively. 

\begin{table}
\centering
\caption{Richness bins for the stacked $\ds(R)$ measurements. The average mass and
bias of clusters in each bin are computed from the best--fit MaxBCG$+$WMAP7 joint model.}
\begin{tabular}{cccc}
\hline
Richness & No. of Clusters $N_j$ & $\avg{\M_{200m}}_j$ [$\hmsun$] & $\avg{b}_j $ \\
\hline
12-17 & 5651 &   1.166$\times 10^{14}$  & 2.511 \\  
18-25 & 2269 &   1.860$\times 10^{14}$  & 3.010 \\  
26-40 & 1021 &   2.918$\times 10^{14}$  & 3.641 \\  
41-70 & 353  &   4.822$\times 10^{14}$  & 4.591 \\  
71+   & 55   &   8.459$\times 10^{14}$  & 6.093 \\  
\hline
\end{tabular}
\label{tab:dsrbins} 
\end{table}

Note that we refer to $\avg{\ln\rich|\M}$ as the mean ``richness--mass''
relation, as distinct from the mean ``mass--richness'' relation
$\avg{\ln\M|\rich}$.  In the presence of scatter, one cannot trivially convert
from one to the other, since there are more low mass halos to scatter to high
richness than vice versa.  Furthermore, since we assume log--normal rather than
Gaussian scatter in richness, the mean richness at a fixed mass $\avg{\rich|\M}$
is {\it not} simply $\exp\avg{\ln\rich|\M}$.  \cite{rozo2012} discussed these
issues in the more general context of cluster observables.  Finally, we caution
that when integrating over a bin in richness, the mean $\ln\M$ is not simply the
value of $\avg{\ln\M|\rich}$ evaluated at the bin center~(e.g., compare centers
of distributions of the same color in the top and bottom panels of
Fig.~\ref{fig:weight}, which we will discuss more later).  These effects account
for the difference between the mean mass--richness relation quoted
by~\cite{rozo2009}, which is directly the inferred mean mass of clusters in
specified bins of richness, and the richness--mass relation of R10, which is a
central power--law derived from a full cosmological fit and more closely
analogous to what we do here.

To account for residual uncertainties in the photometric redshift distribution
used in the weak lensing analysis, we divide the predicted $\ds(R)$ by a
nuisance parameter $\beta$ before comparing with the data, so that we are
effectively modeling the underlying tangential shear profiles while using
$\beta$ to characterize the multiplicative bias in the conversion to $\ds(R)$.
We adopt a Gaussian prior with central value $\be=1.0$ and width
$\delta\beta=0.06$, somewhat larger than the uncertainty of $0.04$ estimated
by~\cite{rozo2009}. We comment more on the constraints on $\beta$ in
\S~\ref{sec:nuisance} and the prior on $\beta$ in \S~\ref{sec:systematics}.

We thus have seven parameters in the model that we fit to the cluster abundance
and large scale $\ds(R)$ data: two cosmological parameters ($\om, \s8$) that we
are hoping to constrain, and five nuisance parameters, among which are ($\la$,
$\lb$, $\sc$) of the richness--mass relation, $\be$ as the residual bias of
the weak lensing shape measurement~(hereafter referred to as the ``weak lensing
bias''), and $\dns$ for the modulation of the $\pks$.

\subsection{Likelihood}
\label{sec:like}

We model the number counts and weak lensing measurements for clusters in
different bins of richness. Aiming for better statistical rather than extra
tomographic constraints, we do not divide our sample into multiple redshift
bins, but only retain the whole photometric redshift range as a single
bin~($0.1<\zph<0.3$). The observable vector in our likelihood model thus has
three components:
\begin{itemize}
\item[1.] $N_i$: number of clusters in each richness bin $i$
for $i \in \{1\cdots9\}$.
\item[2.] $\ds_j(R_k)$: stacked $\ds$ profile of richness bin $j$ 
measured at radius $R_k$, for $j \in \{1\cdots5\}$ and $k \in \{1\cdots6\}$.
\item[3.] Number of clusters with $\rich>120$.
\end{itemize}
We model the combinatorial vector of the $1$st and the $2$nd components as a
multivariate Gaussian~($39$ variables in total), which is fully specified by its
mean vector and covariance matrix.  Fig.~\ref{fig:model_demo} illustrates the
observations of $N_i$~(gray/red histograms in the left panel) and
$\ds_j(R_k)$~(solid circles in the middle panel) used in our analysis. The
richness bins we employed for cluster abundance and weak lensing measurements
are listed in Table \ref{tab:abundancebins} and \ref{tab:dsrbins}, respectively.
We adopt the same richness bins that R10 used for number counts $N_i$ and weak
lensing profiles $\ds_j$; note that the $i$ and $j$ bins are overlapping but not
identical, as larger bins are required to achieve reasonable S/N in $\ds(R)$.
We only use the $\ds$ measurements at large scales~($R_k>5\hmpc$). We comment on
the choice of cutoff radius at $\Rmin = 5\hmpc$ in \S~\ref{sec:xihm}.

For $\rich>120$, the assumption of Gaussian fluctuations in number counts is
invalid due to the rarity of extreme clusters. Following R10, we
model the count of $\rich>120$ clusters as a Poisson binomial distribution,
which is a sum of independent Bernoulli distributions at each integer
$\rich>120$. The likelihood associated with this tail
population of clusters $\ltail$ is given in equation 3 of
\cite{rozo2010}\footnote{Note they have a typo in equation 3 where the order of
the two rhs terms should be reversed}.

The final likelihood $\mathcal{L}$ is then simply the product of the Gaussian
likelihood and the Poisson binomial likelihood.

\subsubsection{Expectation Values}
\label{sec:expval}

For any given cosmology, the mass function $dn/d\M$ and the $\ds(R)$ of DM halos
can be theoretically predicted as functions of mass $\M$ at each redshift $z$.
To convert to the total number $N$ and the average $\ds(R)$ of clusters with
richness $\rich$ and photometric redshift $\zph$, we need to convolve with a
kernel that relates the observables~($\rich$ and $\zph$) to the intrinsic
properties~($\M$ and $z$). For our purpose, the kernel function $\wgtl$ is
defined as the {\it expected} differential number of clusters with mass $\M$ at
redshift $z$ that fall into richness bin $l$ {\it and} within our photometric
redshift range,
\begin{equation}
\wgtl =\frac{\left<d N_l|M,z\right>}{d\M dz},\quad\mbox{$l \in \{i, j\}$}.
\label{eqn:wgt}
\end{equation}
(we use $i$ for number counts, $j$ for $\ds$ bins, and $l$ generic.)
The derivation of $\wgtl$ is similar to that in ~\cite{rozo2007, rozo2010}, and
we briefly describe the procedures below.

\begin{figure}
\centering 
\resizebox{0.47\textwidth}{!}
{\includegraphics{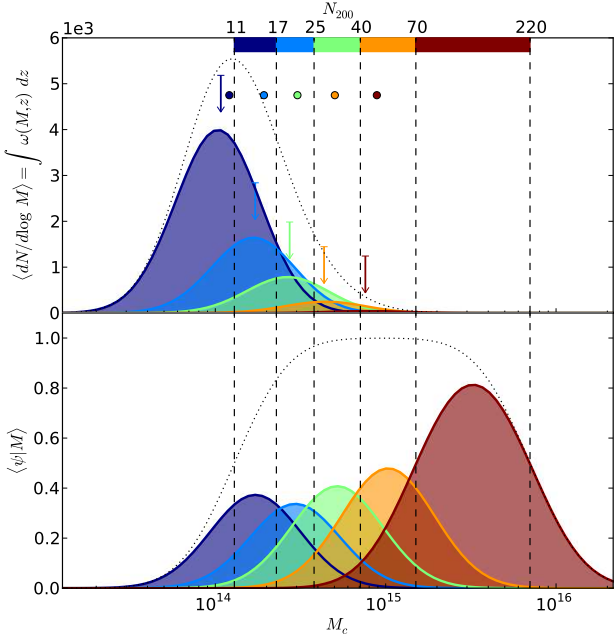}}
\caption{Cluster selection for the five richness bins listed in
Table~\ref{tab:dsrbins}.  
{\it Top}: Distribution of halo masses within each richness bin
$\avg{dN/d\log\M}$, as computed from our best--fit model. The dotted curve gives the
total number distribution of halos within the five bins. Solid circles and
vertical arrows indicate the values of $\avg{\M|\rich}$ and
$\exp\avg{\ln\M|\rich}$, respectively, for each richness bin. 
{\it Bottom}: Richness selection function $\frbinm$. Each
curve shows the probability of halos being assigned to each richness bin at mass
$\M_c$, as computed from our best--fit $P(\rich|\M)$. Dotted curve is the sum of
the five curves, giving the probability of halos being included in the $\ds(R)$
measurements.
In each panel,
each color represents one of the five richness bins. The top x--axis indicates the
mean richness that corresponds to the mass at the bottom x--axis, given by our
best--fit mean richness--mass relation.  Vertical dashed lines show the
demarcation of richness bins based on the top x--axis.}
\label{fig:weight}
\end{figure}

We start by defining the richness selection function $\frbinl$, which is the
probability for a cluster of $\rich$ to be selected into the $l$-th richness
bin. For the richness bins defined by Table~\ref{tab:abundancebins}
and~\ref{tab:dsrbins}, $\frbinl$ is simply a top--hat bracketed by two ends of
the $l$-th richness bin.  For our calculation, however, we need $\frbinml$, the
{\it expected} probability for a cluster of true mass $\M$ to be selected into
the $l$-th richness bin. Without loss of generality, we drop the subscript $l$
of $\psi$ so that
\begin{equation}
\frbinm = \int d\rich P(\rich|\M)\frbin,
\end{equation}
where $P(\rich|\M)$ is the probability of a halo of mass $\M$ observed with
richness $\rich$, specified by the richness--mass relation. The bottom panel of
Fig.~\ref{fig:weight} shows the $\frbinm$ for five
richness bins used for the $\ds(R)$ measurements, given our best--fit model. To
a first approximation, these distributions are Gaussian in $\ln\M$, centered on
the value of $\M$ that corresponds to the mean mass--richness relation at the
central richness of the bin.

We also consider the photometric redshift selection function $\fzphbin$, which
is defined as the probability for a cluster of measured $\zph$ to be selected
into the catalog, i.e., a top--hat bracketed by the redshift extent of the
catalog. Similarly, we instead need the {\it expected} spectroscopic redshift
selection function for a cluster at spectroscopic redshift $z$ to appear in the
catalog
\begin{equation}
\fzphbinz = \int d\zph P(\zph|z)\fzphbin,
\end{equation}
where $P(\zph|z)$ is the probability for a cluster at true redshift $z$ to be
estimated with photometric redshift $\zph$. This probability is assumed to be
Gaussian, centered on spectroscopic redshift $z$, with
$\sigma(\zph|z)=0.008$~\citep{koester2007}.

The product of $\frbinm$ and $\fzphbinz$ then gives the joint selection
probability for a cluster with mass $\M$ at redshift $z$ to be selected into a
given richness bin with $0.1<\zph<0.3$. To predict number density weighted
averages for cluster properties in the bin, we also need the differential
number of clusters with mass $\M$ at redshift $z$, which is simply the product
of the halo mass function $dn/d\M$ and the differential co--moving volume
element $dV/dz$. The final kernel function $\wgt$ is then obtained via
equation~(\ref{eqn:wgt})
\begin{equation}
\wgt = \frac{dn}{d\M}\frac{dV}{dz}\fzphbinz\frbinm,
\end{equation}
the integration of which gives the expectation value for the number counts $N_i$
and $N_j$
\begin{equation}
\langle N_{l} \rangle  = \int d\M dz\,\wgtl,\quad\mbox{$l \in \{i, j\}$}.
\end{equation}
The top panel in Fig.~\ref{fig:weight} shows $\avg {dN/d\log\M}$, the
distribution of halos within richness bins, which is simply the integration of
$\wgt$ over redshift, using our best--fit model parameters. Here we clearly see
the impact of scatter and mass function slope discussed in \S\ref{sec:par}: the
average richness of clusters in a bin of mass is offset from the value of the
mean relation evaluated at the center of the mass bin. For example, a
$10^{15}\hmsun$ cluster would have a high probability~($\sim 40\%$) of being
assigned to the $\rich\in[40-70]$ richness bin~(orange--colored distributions),
but because $10^{15}\hmsun$ clusters are much rarer than lower mass clusters,
the number of them represented in this bin is negligibly small.  Conversely, the
average mass in a bin of richness is offset from the center of $\avg{\psi|\M}$
for that bin --- the average mass of clusters with $\rich\in[40-70]$ is
$\simeq5\times10^{14}\hmsun$~(solid orange circle), and as mentioned in
\S~\ref{sec:par}, the quantity $\exp\avg{\ln\M|\rich}$ is even more offset from
the center of $\avg{\psi|\M}$, landing at $\M\simeq
4.3\times10^{14}\hmsun$~(vertical orange arrow). 

The expectation value for $\ds_j$ is
\begin{equation}
\langle \ds_j \rangle = \frac{1}{\langle N_j \rangle}\int d\M
dz\,\wgtj\,\ds(R|\M,z),
\end{equation}
where $\ds(R|\M,z)$ is the $\ds(R)$ of halos with mass $\M$ at redshift $z$.

For $\ds(R|\M,z)$, by definition~\citep{miralda-escude1991, sheldon2004}, 
\begin{equation}
\ds(R|\M,z) \equiv \bar{\Sigma}(<R|\M,z) - \bar{\Sigma}(R|\M,z),
\end{equation}
where
\begin{eqnarray}
\bar{\Sigma}(<R|\M,z)&=& \int_0^{R}dr_p r_p \int_{-\infty}^{+\infty}dr_z\xihm\left(\sqrt{r_p^2+r_z^2}, \M\right)
\nonumber\\
&\times&\rho_{\mrm{m},z}\frac{2}{R^2} 
\label{eqn:dsr}
\end{eqnarray}
and
\begin{equation}
\bar{\Sigma}(R|\M,z) = \rho_{\mrm{m},z}\int_{-\infty}^{+\infty}\xihm\left(\sqrt{R^2+r_z^2} ,
\M\right)dr_z.
\end{equation}
In the above equations, $\rho_{\mrm{m},z}=\om\rho_{\mrm{c},0}(1+z)^3$ is the mean density of the universe
at $z$, and $\xihm(r, \M)$ is the halo-matter cross-correlation function at
3-D distance $r$ for halos with mass $\M$. 

For $\xihm(r, \M)$, we use a variant of the model proposed by HW08 
\begin{eqnarray}
\xihm(r, \M) &=&   \left\{
\begin{array}{ll}
 \xi_\mrm{1h} & \quad\mbox{if $\xi_\mrm{1h} \geqslant \xi_\mrm{2h}$ },\nonumber\\
 \xi_\mrm{2h} & \quad\mbox{if $\xi_\mrm{1h} < \xi_\mrm{2h}$ },\nonumber
\end{array}
\right.\\
\xi_\mrm{1h} &=& \frac{\rho_\mrm{halo}(r, \M)}{\rho_\mrm{m}} - 1, \nonumber\\
\xi_\mrm{2h} &=& b(\M) \; \xinl.
\label{eqn:xihm}
\end{eqnarray}
Here $\xi_\mrm{1h}$ and $\xi_\mrm{2h}$ are the so-called ``1-halo'' and
``2-halo'' terms in the halo model~(see \citeauthor{cooray2002} 2002 for a
review), $\rho_\mrm{halo}(r, \M)$ is the NFW density profile of halos with mass
$\M$, and $b(\M)$ is the halo bias function. The difference from the original
HW08 prescription is that we use the non-linear matter autocorrelation $\xinl$
computed from the fitting formula of~\cite{smith2003} instead of the linear
prediction $\xilin$. We demonstrate that this modification provides an accurate
approximation of large scale measurement of $\ds(R|\M,z)$ from N-body
simulations in \S~\ref{sec:xihm}.

\subsubsection{Covariance Matrix}
\label{sec:covmat}
 
The covariance matrix of the model $\cov$ is comprised of three sub--blocks,
$\covnn$~(abundance--abundance), $\covns$~(abundance--shear), and
$\covss$~(shear--shear). We begin with $\covnn$, which has two independent
sources of uncertainties: 
\begin{itemize}
\item Sample variance due to limited survey volume~(a.k.a. cosmic variance). 
\item Poisson fluctuations in cluster number counting~(a.k.a. shot noise).
\end{itemize}
Thus the covariance between cluster number counts in two
richness bins $i$ and $i'$ is the sum of two components
\begin{equation}
\covnnii = \covnnii^\mrm{sample} + \covnnii^\mrm{Poisson},
\end{equation}
and, as will become apparent below, the diagonal parts of both
components scale with survey volume in a similar fashion. We will describe each
in turn. 
 
Assuming the clustering bias $b$ of clusters is linear with respect to the
underlying density fluctuation on the scale of the survey $R_\mrm{V}$, the
sample variance term is simply~\citep{hu2003}
\begin{equation}
\covnnii^\mrm{sample} = \langle N_i \rangle \langle N_{i'} \rangle \langle b_i
\rangle \langle b_{i'} \rangle  \sigma^2(R_\mrm{V}),
\label{eqn:cnnsam}
\end{equation}
where
\begin{equation}
\langle b_{l} \rangle = \frac{1}{\langle N_{i,i'} \rangle}\int
d\M dz\,b(M)\,\wgtl,\quad\mbox{$l \in \{i, i'\}$},
\end{equation}
and $\sigma^2(R_\mrm{V})$ is the variance of the linear density fluctuation
field on scale $R_\mrm{V}$, which we assumed to be adequately approximated by
the radius of a sphere that has the same volume as the survey. On relevant
scales, $\sigma^2(R)$ is approximately proportional to $1/V$, so the diagonal
terms $\covnniidiag^\mrm{sample} \propto \langle N_i^2 \rangle/V \propto V
$~\citep{hu2003}.

The Poisson fluctuation term is trivial, 
\begin{equation}
\covnnii^\mrm{Poisson} = \delta_{ii'} \langle N_i \rangle,
\label{eqn:cnnpoi}
\end{equation}
i.e., it is diagonal and the variance is equal to the expectation value of the
number counts in each richness bin.  Our covariance matrix differs from that of
R10 in that we do not include the ``stochasticity'' contribution in R10. One
might think that scatter between richness and mass would introduce off--diagonal
covariances because clusters that scatter out of one richness bin will scatter
into a neighboring bin.  Indeed, when the total number of halos is held fixed,
such covariance do occur.  However, when one simultaneously considers both
Poisson sampling and the stochasticity due to scatter in the richness--mass
relation, one finds that the two are coupled in such a way that the naive
Poisson terms represent the full covariance matrix. We have explicitly verified
this via numerical experiments.  Simply adding the Poisson and ``stochastic''
covariance matrices, as was done in R10, is incorrect, and the naive Poisson
term alone captures the full variance due to Poisson fluctuations and the
stochasticity of the richness--mass relation.

%One might think that
%scatter between richness and mass would introduce off--diagonal covariances
%because clusters that scatter out of one richness bin scatter into a neighboring
%bin. However, once one accounts for Poisson fluctuations in the number of
%clusters at a given mass, the Poisson contribution of Equation~\ref{eqn:cnnpoi}
%turns out to be complete, a point that we verified for ourselves explicitly with
%numerical experiments. The discussion of R10 incorrectly suggests otherwise;
%Specifically, there is no ``stochasticity'' contribution of off--diagonal terms
%to $\covnnii$.

\begin{figure}
\centering
\resizebox{0.47\textwidth}{!}
{\includegraphics{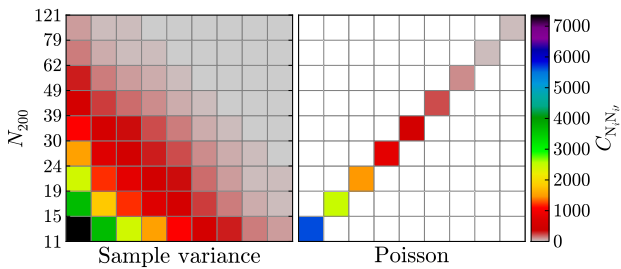}}
\caption{Comparison between the two components of $\covnnii$:
$\covnnii^\mrm{sample}$~(left) and $\covnnii^\mrm{Poisson}$~(right), as
computed for the best--fit model. Each panel is a $9\times9$ matrix color-coded
by the value of $\covnnii$. The richness bins are shown on the y-axis of the
left panel.}
\label{fig:covnn_demo}
\end{figure}

Fig.~\ref{fig:covnn_demo} compares the two components of $\covnnii$ as computed
for our best--fit model. The diagonal elements of the sample variance component
are much weaker than the Poisson errors except for the lowest richness bin,
where sample variance dominates~\citep{hu2003}. Sample variance also produces
positive off--diagonal terms. Overall, the off--diagonal terms are smaller than
diagonal terms, but not completely negligible.

The other two sub--blocks, $\covns$ and $\covss$, also have contributions from
sample variance in the cluster--mass correlation function. However, the dominant
contributions to the $\ds(R)$ errors come from statistical errors in the weak
lensing measurements themselves. One major contribution to these errors is shape
noise from the random orientations of source galaxies; with rms galaxy
ellipticity $\sim 0.3$, one needs $N\sim(0.3/\gamma)^2$ sources to measure a
shear $\gamma$ at $S/N=1$, and with the surface density $n_\mrm{eff}\sim 1
\mrm{arcmin}^{-2}$ typical of SDSS weak lensing data, the $S/N\ll1 $ for any
individual clusters. (Tangential shear is roughly $\gamma_T\sim 0.01$ near the
cluster virial radius and smaller beyond.) A second contribution comes from
coherent cosmic shear lensing of source galaxies by foreground or background
structure not associated with the lensing cluster~\citep{dodelson2004, hoekstra2011-1}. For the $\ds(R)$
covariance matrix in each richness bin, we use the empirical estimates of S09
based on jackknife re--sampling of the data set in large area patches, where the
measurement regions around clusters do not overlap, shape noise errors should be
diagonal and drop with the square root of the number of source galaxies in each
radial bin. However, as mentioned in S09, the jackknife errors at $R>5\hmpc$
become substantially larger than this naive expectation, and there are
significant off--diagonal terms for different radial bins. These large,
correlated errors presumably reflect the coherent cosmic shear effect described
above, though they could also be affected by spatially coherent fluctuations in
the quality of point spread function~(PSF) correlation. In principle there
is also a $\covns$ sub--block to the covariance matrix, but we ignore it here
because the $\ds(R)$ errors are dominated by measurement error not sample
variance.

Additionally, the entire covariance matrix is affected by uncertainties in the
completeness and the purity of the MaxBCG cluster sample. Interlopers and
missing clusters can both bias the number counts, though the two effects tend to
cancel.  For $\ds(R)$,  missing clusters only increase errors, while interlopers
can bias the measurement via dilution. The magnitude of this effect is difficult
to estimate without detailed simulations, since the main source of
``interlopers'' will be random superpositions of smaller clusters at different
redshifts, and the weak lensing signal from the superposed systems may be
similar to that expected from a single system of the estimated richness~(in
which case there is no dilution).  For $\rich>11$, the purity and completeness
of the MaxBCG catalog is estimated to be at the $\sim95\%$ level or higher, so
any associated biases should be small, though they can be coherent across
richness and radial bins. Following R10, we define the magnitude of this bias to
be $\lambda\simeq 1\pm0.05$ and add $\mrm{Var}(\lambda)=0.05^2$ to the fractional
errors in all elements of $\covnn$ and diagonal elements of $\covss$,
respectively. We comment on our treatments for $\lambda$
in~\S~\ref{sec:systematics}.

\subsection{Priors} 
\label{sec:priors}
\begin{table}
\centering
\caption{Prior Specifications. The $\pks$ is generated from the best--fit
cosmological parameters from Reid10, and $\scm$ is not a model parameter but an
observable. Priors that contain the form $[a, b]$ mean the parameter in
question is restricted to values within that range.  Priors that contain the
form $x=a\pm \delta a$ refer to a Gaussian prior of mean $\avg{x}=a$ and
variance $\mathrm{Var}(x)=(\delta a)^2$. The combination of the two forms is a
Truncated Gaussian. Uninformative priors mean the parameter in question is
absolutely unrestricted.}
\begin{tabular}{cc}
\hline
Parameter & Prior \\
\hline
$\om$  & Uniform on [0.05, 0.95]     \\
$\s8$  & Uniform on [0.40, 1.20]     \\
$\sc$  & Uniform on [0.10, 1.50]     \\
$\be$  & Truncated Gaussian $1.00\pm 0.06$ on [0.50, 1.50] \\
$\la$  & Uninformative               \\
$\lb$  & Uninformative               \\
$\dns$ & Truncated Gaussian $0.000\pm 0.013$ on [-0.1, +0.1] \\ 
\hline\\
$\pks$ & ``$\lcdm$'' Column of table~3 in \cite{reid2010} \\
$\scm$ & Gaussian $0.45\pm 0.10$; \cite{rozo2009} \\
\hline
\end{tabular}
\label{tab:priors} 
\end{table}

Table~\ref{tab:priors} summarizes the priors assumed in our analysis. To ensure
that our results are driven by the data, we place either uniform or
unrestrictive priors on five of our seven model parameters, $\om$, $\s8$, $\sc$,
$\la$, and $\lb$, and conservative truncated Gaussian priors on the other two,
$\beta$ and $\dns$. We take the $\pks$ to be that of the best--fit $\lcdm$ model
from Reid10, multiplied by $(k/1\,h\mathrm{Mpc}^{-1})^{\dns}$ to allow minor
modulation.~\footnote{The choice of pivotal wavenumber is arbitrary as it does
not affect the $\pks$.} Note that while this is a ``$\lcdm$'' power spectrum for
specific cosmological parameters, we are treating it here as the empirical
description of the observed shape of the galaxy power spectrum.
We also place priors on $\scm$, the converse scatter
defined as the dispersion in log--mass in the fixed richness bin $\rich$=[38,
42]. Following R10, we take the prior on $\scm$ directly from the analysis in
~\cite{rozo2009}, $\scm=0.45\pm0.10$, which is derived by requiring consistency
among the MaxBCG $L_X$--$\rich$ relation, the MaxBCG richness--mass relation
from weak lensing, and the $L_X$--$\M$ relation measured in the 400d
survey~\citep{burenin2007} by \cite{vikhlinin2009}, as well as the scatters
within each of the three scaling relationships.

\section{Tests using Simulations and  Mock Data Sets}
\label{sec:simumock}

\begin{figure*}
\centering
\resizebox{0.90\textwidth}{!}
{\includegraphics{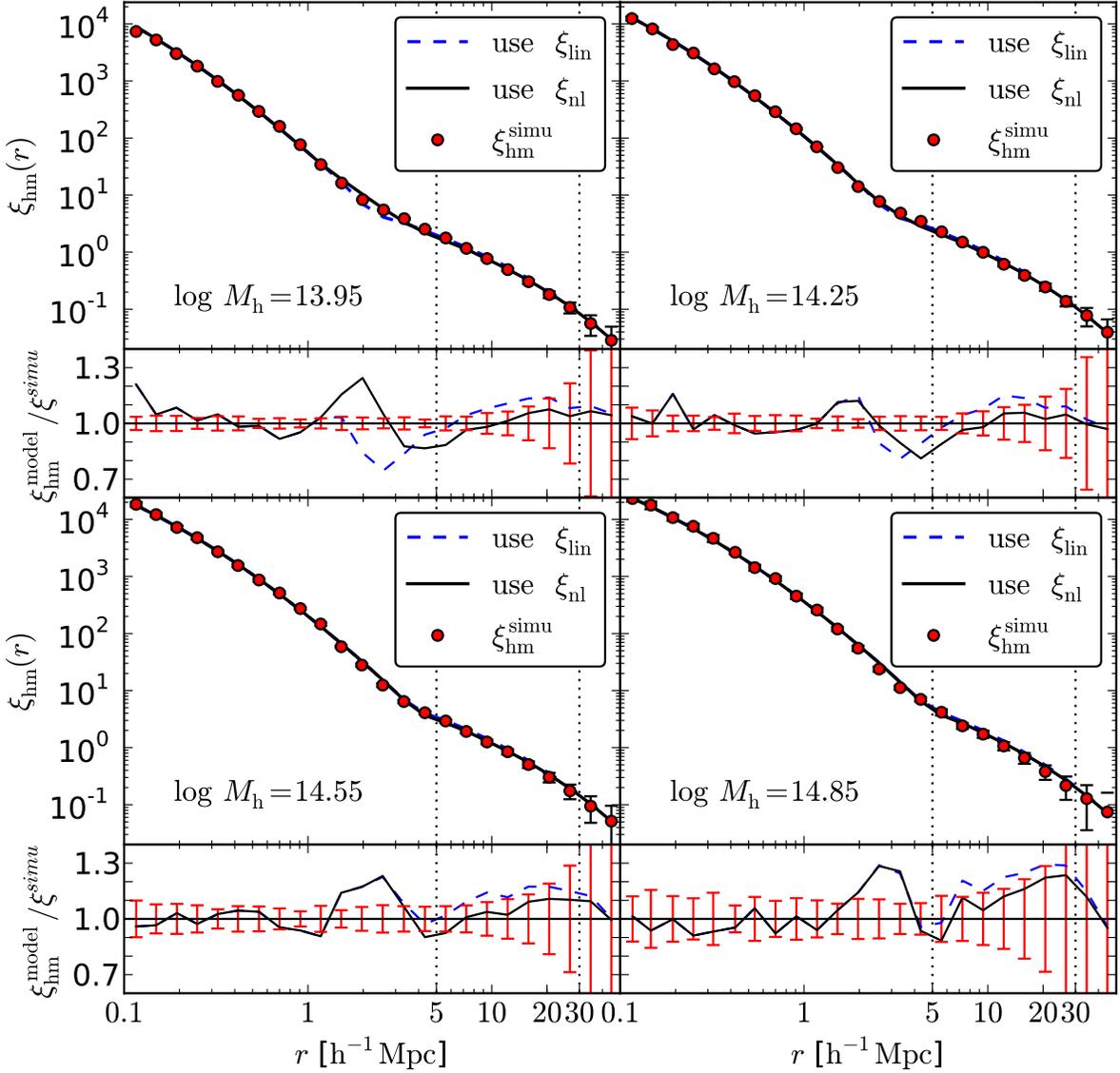}}
\caption{
Comparison between the 3D halo--matter cross--correlation profiles predicted by
the original HW08 prescription~(dashed lines), predicted by our modified
prescription that uses the non-linear correlation function~(solid lines), and measured from simulation~(solid circles with
errorbars), for four different halo masses at $z=0$. For each mass, the lower
subpanel shows the ratios between model predictions and simulation measurement.
Errorbars on the simulation measurement in the upper panels are estimated from
jackknife sub--sampling of the simulation box and propagated to the ratio
curves. Our analysis in this paper uses measurements of $\ds(R)$ beyond a
projected co-moving separation of $5\hmpc$, marked by the vertical dotted lines.
}
\label{fig:xihm}
\end{figure*}

\subsection{Basic Implementation}

For the linear matter power spectrum in our likelihood calculation, we take the
cosmological parameters inferred by Reid10 and compute $\plin$ using the
low--baryon transfer function of~\cite{eisenstein1999}, which is a good
approximation to the full transfer function on scales well below the BAO scale.
We refer to this linear power spectrum as $\preid$, as it fits their power
spectrum measurements of SDSS galaxies by construction.  To compute the
non-linear matter correlation function $\xinl$, we use the prescription from
~\cite{smith2003} to generate $\pnl$ for Fourier transforming to $\xinl$. The
halo mass function and the halo bias function are from ~\cite{tinker2008} and
~\cite{tinker2010}, respectively, with the halo mass defined by $\M \equiv
\M_{200m} =  200 \rho_m V_\mrm{sphere}(r_{200m})$.  We use the
NFW~\citep{navarro1996} halo density profile for $\rho_\mrm{halo}(r, \M)$. The
halo mass--concentration relationship is from the fitting formula of
~\cite{zhao2009}, which accurately recovers the flattening of halo concentration
at high masses. (In MaxBCG we do not expect an upturn, which only shows up at
redshifts beyond 1; see~\citeauthor{prada2012} 2012). In the parameter inference stage, the posterior
distribution is derived using a Markov Chain Monte Carlo~(MCMC), where an
Adaptive Metropolis step method is utilized during the burn--in period to expedite
the exploration of highly correlated parameter space. For each MCMC chain, we
perform $320,000$ iterations, $20,000$ of which belong to the burn-in period for
adaptively tuning the steps. To eliminate the tiny amount of residual
correlation between adjacent iterations, we further thin the chain by a factor
of $10$ to obtain our final results.

\subsection{Halo-Mass Correlation Function}
\label{sec:xihm}
\begin{figure*}
\centering
\resizebox{0.9\textwidth}{!}
{\includegraphics{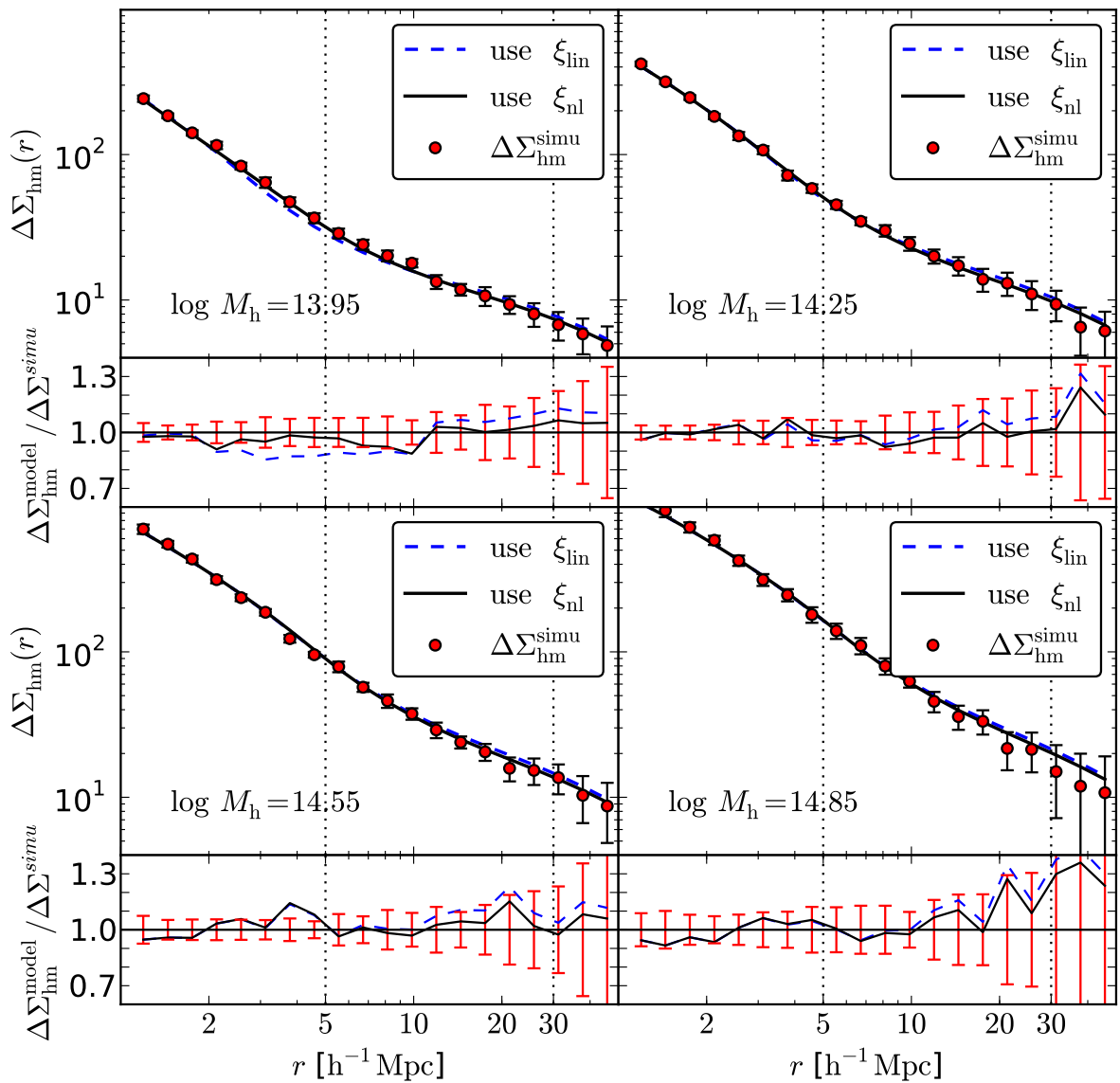}}
\caption{Similar to Fig.~\ref{fig:xihm} but for the 2D halo surface density
contrast profiles. 
}
\label{fig:dshm}
\end{figure*}

To extract maximum cosmological information from the S09 $\ds(R)$ measurements,
one should simultaneously fit the data on all scales. However, as already
discussed in the introduction, we have elected to focus on large scales in this
paper so that our constraints are complementary to those derived for the MaxBCG
sample by R10, who measured the cluster mass function using a richness--mass
relation calibrated by the S09 data on small scales.  Restricting our analysis
to large scales also allows us to avoid two sources of systematic error, one
observational and one theoretical. The observational systematic is the effect of
cluster mis--centering, which tends to depress $\ds(R)$ at small scales.  This
effect can be estimated from detailed simulations, but with some uncertainties
associated with the baryonic physics~\citep{sanderson2009} and the optical
cluster finder~\citep{johnston2007}. The theoretical systematic is the
uncertainty in the halo--mass correlation function in the transitional region
between the NFW halo mass profile and the large--scale regime where it is a
linearly biased multiple of the matter correlation function. We choose our
minimum scale $\Rmin$ so that this theoretical systematic is small compared to
the observational uncertainties of the S09 measurements.

\begin{figure*}
\centering
\resizebox{1.0\textwidth}{!}
{\includegraphics{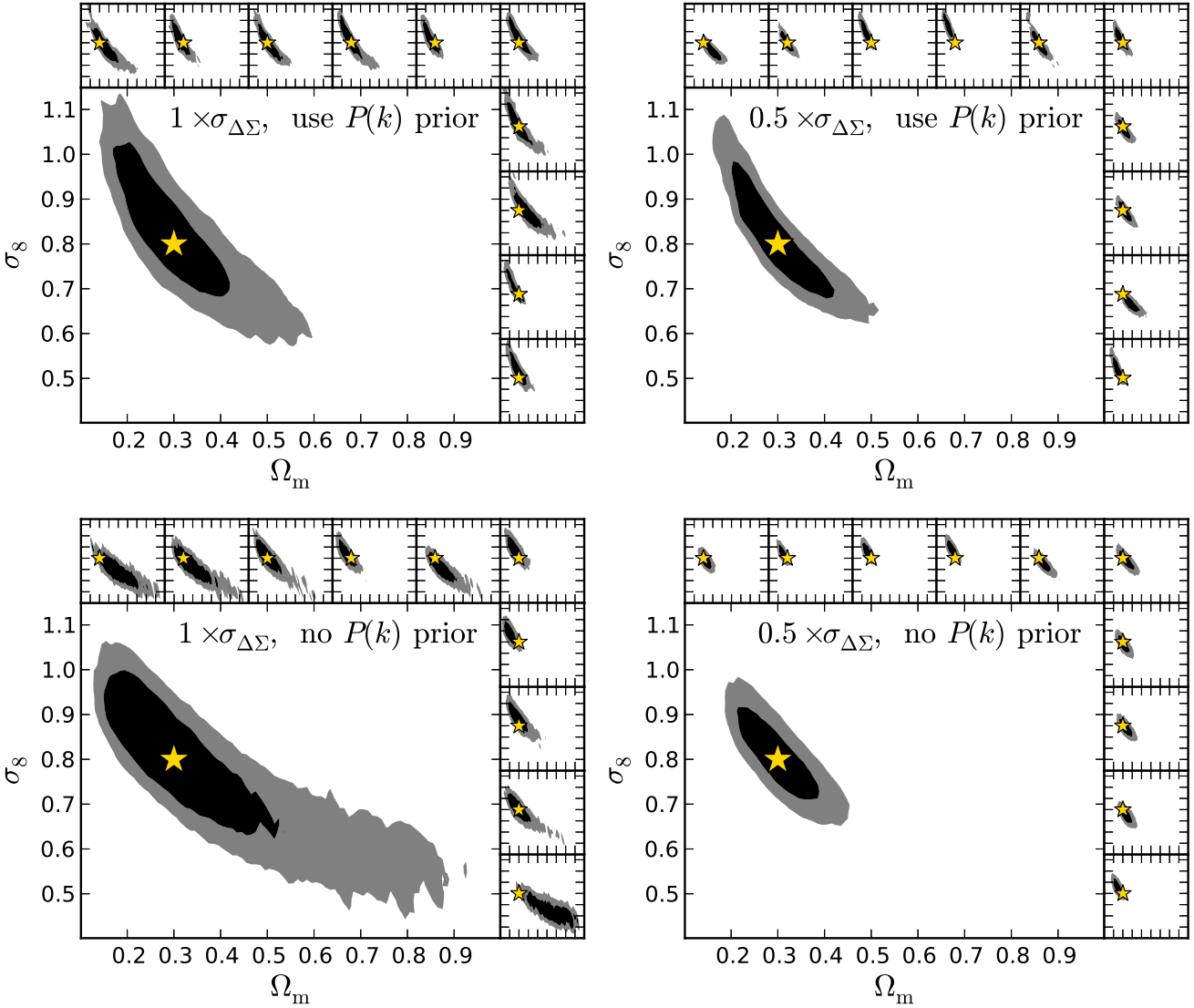}}
\caption{
The effect of the $\pks$ prior on the $\om$--$\s8$ constraint using mock data.
The two left panels compare the confidence regions derived with~(top) and
without~(bottom) using the $\pks$ prior for the mock data with the original
jackknife errors. The two right panels show a similar comparison for the mock
data with weak--lensing errors shrunk by $1/2$.  In each panel, the contours
in the embedded sub-panels are the confidence regions for the 10 individual
realizations, and the contours in the main panel are the results after combining 
all 10 MCMC chains.  Each set of contours shows $95\%$ and $68\%$ confidence
regions inwards, and the star on top indicates the input cosmology~($\om$=0.30,
$\s8$=0.80) for generating the mock data. For the ``no $P(k)$ prior'' cases, we
compute the linear power spectrum explicitly for each iteration in the MCMC
chains.
}
\label{fig:pkshape}
\end{figure*}

To determine the cutoff scale $\Rmin$, and to test the accuracy of our model of
$\ds(R)$ beyond $\Rmin$, we use halos in cosmological simulations.  
The simulation we use is the ``L1000W'' presented in ~\cite{tinker2008,
tinker2010}, where the halo mass function and bias function are calibrated. It
evolves $1024^3$ particles with $m_p = 6.98\times10^{10}\hmsun$ in a periodic box of
co--moving length $1000 \hmpc$ using the Adaptive Refinement
Tree~\citep[ART;][]{kravtsov1997, gottloeber2008} code. For the mass scales we consider here,
$\M>5\times10^{13}\hmsun$, it has at least $\sim 1000$ particles for each halo
and is thus well--suited for the study of $\xihm$ and $\ds$.  For the model of
$\xihm$, we adapt the original HW08 prescription by using $\xinl(r)$ as in
equation~(\ref{eqn:xihm}), so that $\xihm$ is more accurate on large scales.
There are other formulas for $\xihm$ in the literature, but it is not 
clear whether any formulation applies universally across cosmologies. The large
scale behavior should be dictated by linear theory in any case, and the HW08
model appears adequate for our present application.

Fig.~\ref{fig:xihm} compares different $\xihm$ profiles predicted by the
original HW08 prescription~(dashed lines), predicted by our modified
prescription~(solid lines), and measured from simulation~(solid circles) at
$z=0$, for four halo masses that span the relevant mass range of our cluster
sample. The errorbars are from jackknife re--sampling of octants of the
simulation box.  Both the original and the modified models recover $\xihm$ on
small scales ($<1\hmpc$) very well~(within $10\%$) because of the success of the
NFW profile in describing halo density profiles within $r_\mrm{vir}$. On the
transitional scales between $1\hmpc$ and $5\hmpc$, both models show
discrepancies with the simulation of up to $20-30 \%$ due to the 
discontinuous change of the prescription between the 1--halo and 2--halo
regimes. Beyond $5\hmpc$, the modified model clearly outperforms
the original one, agreeing with the simulation within the errorbars in all mass
bins, and agreeing to within $10\%$ except for the highest mass bin, where the
measurement error is large due to the small number of very massive clusters in
the simulation. (In detail, the $6\hmpc$ prediction is outside the error bar in
the two lowest mass bins.)

Fig.~\ref{fig:dshm} shows the same comparison between the two models and the
simulation measurement for $\ds(R)$, the quantity we care most about. Similar to
the $\xihm$ case, the models agree with the simulation to within a few per cent
below $\sim 2\hmpc$, and for most of the transitional scales between $2\hmpc$
and $5\hmpc$; projection dilutes but does not eliminate the effect of the
``discontinuity spikes'' seen in Fig.~\ref{fig:xihm}. Beyond $5\hmpc$, the
original HW08 model generally has some $>15\%$ deviations from the simulation
measurement in all mass bins, while the modified model is in excellent agreement
with the simulation, to within $5\%$ in the three lowest mass bins and within
the measurement uncertainties~($30\%$) in the highest mass bin. Note that the
S09 $\ds$ measurement also has $\sim 30\%$ uncertainties on large scales.
We test the effect of dropping the $\ds(R)$ measurements for the
highest richness bin in \S~\ref{sec:systematics}.

To bracket the redshift extent of the MaxBCG clusters, we have also done the
comparison test using the simulation output at $z=0.3$, and the results are
similar.  Based on the results of the tests, we conclude that the impact of
scale-dependent bias on $\ds(R)$ is very weak on scales beyond $5\hmpc$ using
the modified HW08 model in Equation~\ref{eqn:xihm}, well below the uncertainties
in the S09 measurements. Therefore, we choose to use the stacked weak lensing
observations beyond co--moving scale of $\sim 6\hmpc$, which for the redshifts
of our cluster sample corresponds to a cutoff radius~$\Rmin$ of $5\hmpc$ in
physical units.

\subsection{Power Spectrum Shape as a Prior}
\label{sec:testshape}

Once we have determined $\Rmin$, the large scale 3D density contrast profile of
halos, $\om\xihm(r,\M)$, carries clean and easily accessible cosmological
information, as $\om\xihm(r,\M) \propto \om b(\M) \s8^2$ if we know the shape of
the matter correlation function $\ximm(r)$ well. However, our measurements of
$\ds(R)$ provide only limited constraints on the shape of $\ximm(r)$, in part
because of projection, in part because we use only large scale measurements, and
most of all because the statistical errors in the $\ds(R)$ measurements remain
fairly large. Uncertainty in the $\ximm(r)$ shape would limit our ability to
optimally combine measurements from multiple scales, and it would limit our
ability to translate our measurements from these scales to a value of $\s8$,
which is defined at the specific scale of $8\hmpc$.

To circumvent this problem, we introduce the shape of the power spectrum
measured from SDSS Luminous Red Galaxies by Reid10 as a prior in our
Bayesian analysis, without introducing any explicit priors on individual
cosmological parameters. We are taking advantage of the fact that the shape of
the power spectrum is well constrained observationally by galaxy clustering
data, even though it is not well constrained by our $\ds(R)$ measurements. We
allow deviations from $\preid$ as parameterized by $\dns$.  To test the
performance of the $\pks$ as a prior and its sensitivity to uncertainties in
$\ds(R)$, we generate two sets of mock data, one using the original S09
jackknife errors, and one using half the S09 errors. For each mock set, we
produce 10 random realizations from the multivariate Gaussian describing the
cluster counts and $\ds(R)$ values, using the
parameters $\om=0.30$, $\s8=0.80$, $\sc=0.36$, $\be=1.0$, $\la=2.4$, $\lb=4.2$,
and $\dns=0.0$, along with other cosmological parameters set as the WMAP7
values. We also compute the $\pks$ from the same cosmology, and we perform MCMC
analyses on the 10 random realizations of each mock data set with and without
using the $\pks$ prior. When there is no shape prior used, we fix all the
cosmological parameters to be their WMAP7 values except for $\om$ and $\s8$, and
the $\pks$ varies with $\om$ according to the $\lcdm$ prediction.

Fig.~\ref{fig:pkshape} shows the effect of introducing the $\pks$ prior on the
$\om$--$\s8$ constraints for the two mock data sets. For the mock data with
original weak--lensing errors~(left two panels), when the $\pks$ is {\it a
priori} unknown~(bottom left panel), the analysis generally accepts an incorrect
region of high--$\om$ and low--$\s8$ within $95\%$ confidence, allowing models
in which the associated $\pk$ becomes much bluer due to an earlier epoch of
matter--radiation equality and transforms into a $\ximm$ that is too
strong~(weak) on small~(large) scales.  Despite being physically unlikely, after
projection this model leads to $\ds(R)$ profiles that are consistent with the
mock data within the errors. When the shape prior is used~(top left panel), the
high--$\om$ and low--$\s8$ region is correctly rejected by the model,
demonstrating the efficacy of the shape prior in eliminating the degeneracy
between uncertainties in the $\pks$ and cosmological parameters.

For the mock data with smaller weak--lensing errors~(right two panels), both
$\om$ and $\s8$ are well constrained either with or without using the shape
prior, showing that the shape degeneracy greatly diminishes with the reduced
weak-lensing errors even though we do not use small scale $\ds(R)$ information.
The contours are slightly more elongated when the shape prior is used~(top right
panel), because it is easier for extreme values of $\om$ and $\s8$ to fit the
observed cluster richness function when the right $\pks$ is directly known
compared to when the right $\ns$, $\ob$, $\on$, and $h$ are known. In other
words, incorrect values of $\om$ now produce deviations in the shape of $\ds(R)$
that are detectable with the smaller errors. Fig.~\ref{fig:pkshape} also
demonstrates that the uncertainties of our constraints depend crucially on the
statistical errors in the weak lensing measurement on large scales --- the
$1\sigma$ uncertainties in $\om$ and $\s8$ are reduced by $\sim 45\%$ after
$\sigma_{\ds}$ shrinks by a factor of two in the lower right panel.
 
\section{Parameter Constraints}
\label{sec:results}
\begin{table}
\centering
\caption{Best--fit Models}
\begin{tabular}{ccc}
\hline
Parameter & MaxBCG & MaxBCG+WMAP7\\
\hline
$\om$  &   $0.325_{-0.067}^{+0.086}$  &  $0.298_{-0.020}^{+0.019}$    \\
$\s8$  &   $0.828_{-0.097}^{+0.111}$  &  $0.831_{-0.020}^{+0.020}$    \\
$\sc$  &   $0.432_{-0.068}^{+0.063}$  &  $0.436_{-0.024}^{+0.012}$    \\
$\be$  &   $1.004_{-0.060}^{+0.060}$  &  $0.968_{-0.030}^{+0.034}$    \\
$\la$  &   $2.446_{-0.127}^{+0.142}$  &  $2.465_{-0.052}^{+0.094}$    \\
$\lb$  &   $4.148_{-0.229}^{+0.249}$  &  $4.163_{-0.068}^{+0.129}$    \\
$\dns$ &   $0.001_{-0.013}^{+0.013}$  &  $0.001_{-0.001}^{+0.001}$    \\
\hline
\end{tabular}
\label{tab:bestfit} 
\end{table}

\begin{figure*}
\centering
\resizebox{1.0\textwidth}{!}
{\includegraphics{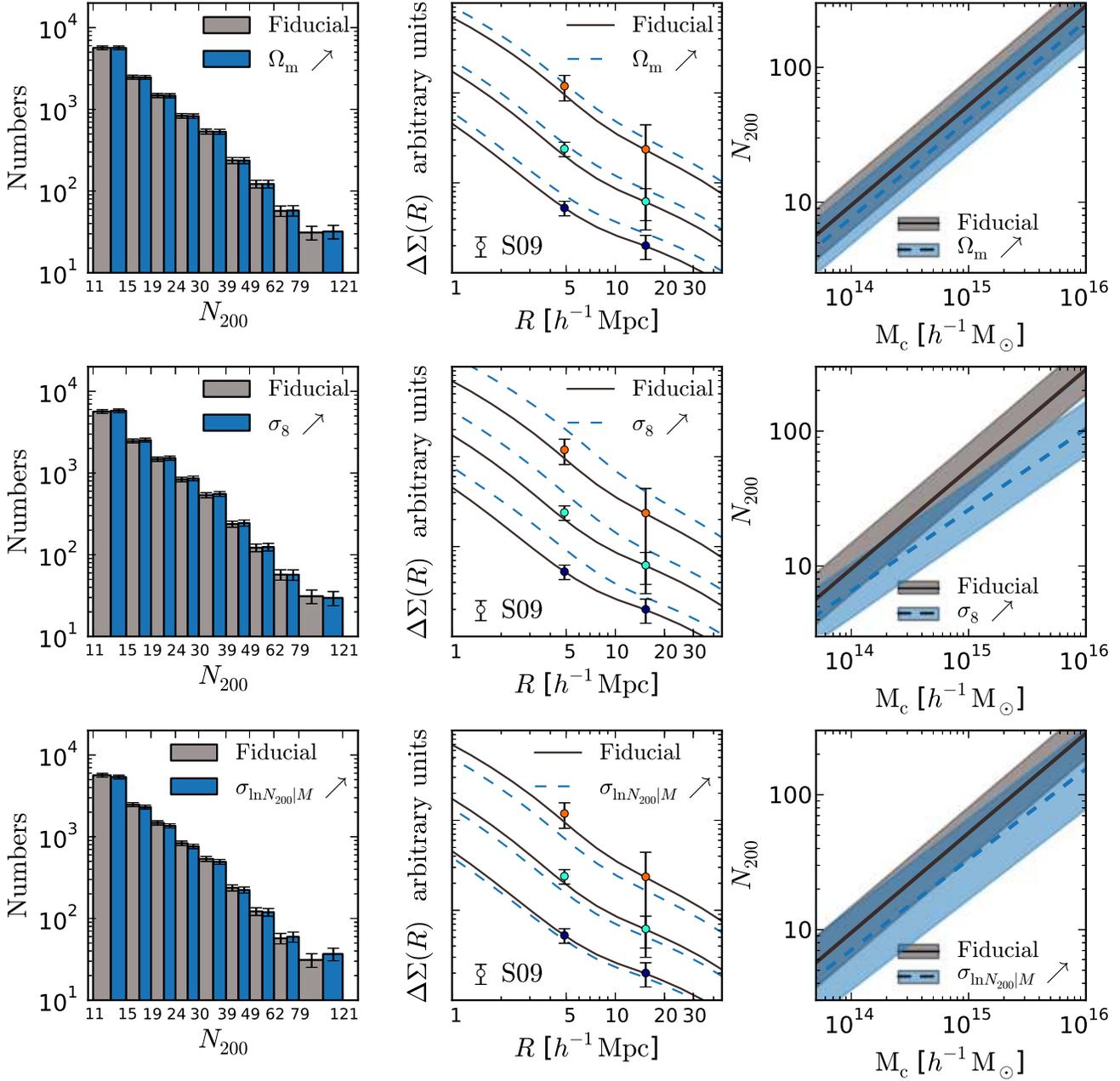}}
\caption{Illustration of the underlying methodology of our analysis. In each
row, one of the three key parameters~(from top to bottom: $\om$, $\s8$, and
$\sc$) is increased from its fiducial value by a factor of $1.5$, and we vary the mean richness--mass
relation~(blue vs. gray bands in the right column) to find a new best--fit
to match the cluster number counts~(blue vs. gray histograms in the left
column). The panels in the middle column then compare the $\ds(R)$ profiles
predicted by the new best--fit model~(blue dashed curves) to the fiducial
profiles~(gray solid curves). Points with error bars indicate the S09
measurements of $\ds(R)$ at $5\hmpc$ and $15\hmpc$ for the three richness bins.
See the text for more details.}
\label{fig:pedagogical}
\end{figure*}
\begin{figure*}
\centering
\resizebox{1.0\textwidth}{!}
{\includegraphics{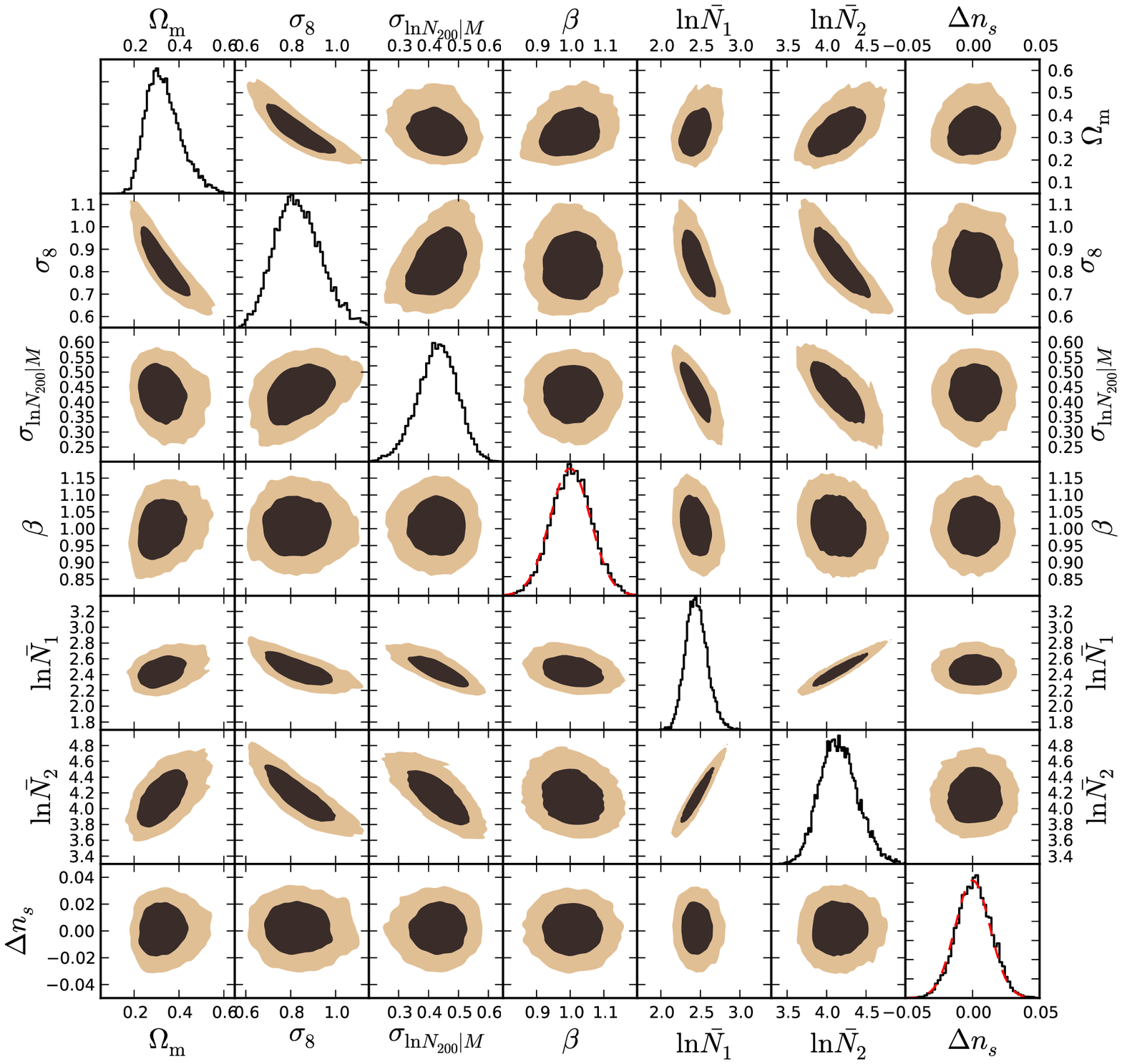}}
\caption{Confidence regions from our analysis of the MaxBCG data in the 2D planes that comprised of all the pair
sets of model parameters. Histograms in the diagonal panels show 1D
posterior distributions of individual parameters. Contour levels run through
confidence limits of $95\%$~(light brown) and
$68\%$~(dark brown) inwards. The assumed prior distributions for $\be$ and $\dns$
are shown as dashed curves on the fourth and the seventh diagonal panel,
respectively; they are barely distinguishable from the posterior distributions.
}
\label{fig:datacontour}
\end{figure*}

Our best--fit model is summarized in the first column of
Table~\ref{tab:bestfit}, where for each parameter we quote the median~($50\%$)
as central value and the $[18.54\%$, $84.16\%]$ interval as $\pm 1\sigma$
uncertainties. 

Before presenting the detailed results of our analysis, it is instructive to
illustrate how the addition of large scale weak lensing measurements helps break
the degeneracy in cluster abundance measurements among the cosmological
parameters $\om$ and $\s8$ and the nuisance parameter $\sc$. The experiments
shown in Fig.~\ref{fig:pedagogical} are designed to serve this purpose,
providing a more detailed form of the approximate argument sketched in the
introduction.  Starting from the fiducial model~(Table~\ref{tab:bestfit}, column
2) that matches both the cluster number counts and $\ds(R)$ data, in each row of
Fig.~\ref{fig:pedagogical}, we raise one of the three key parameters~(from top
to bottom: $\om$, $\s8$, and $\sc$) by a factor of $1.5$ from its fiducial value
while keeping the other two parameters fixed at their fiducial values. We then
re--fit to the cluster abundance data {\it alone} by varying the mean
richness--mass relation. The new best-fit model in each row thus represents one
of three families of false models that are indistinguishable from the underlying
true model if we only employ the cluster abundance data for constraint. With the
modified richness--mass relation in the right column, the new best-fit model
predicts cluster number counts that can match the original data (left column)
but different $\ds(R)$ profiles that cannot~(middle column). In detail,
\begin{itemize}
\item When $\om$ is increased~(top row), there is no change in the amplitude of
clustering except for an overall change of density, so the halo mass and bias
functions both shift uniformly to higher masses. Matching the observed number
counts only requires a decrease in the overall amplitude of the mean
richness--mass relation. Since there is no change in the cluster bias at fixed
richness, the boost in the $\ds(R)$ profiles is directly caused by the increase
in $\om$ and is thus independent of distance and richness.
\item When $\s8$ is increased~(middle row), the halo mass function changes
amplitude and shape: the abundance of halos throughout the cluster mass regime
increases, but the increase is larger at higher masses. To fit the observed
abundance as a function of richness, the mean richness--mass relation must tilt
downward, becoming shallower than the fiducial relation. This change suffices to
reproduce the original number counts, but the $\ds(R)$ profiles shift upward
because of the growth in the amplitude of $\ximm(r)$. This growth~($\propto
\s8^2$) is partly compensated by a reduction in halo bias factors, but this
reduction is mass--dependent, with the consequence that $\ds(R)$ grows more for
high richness clusters than for low richness clusters.
\item When $\sc$ is increased~(bottom row), it has very similar effect on
cluster number counts as increasing $\s8$, scattering progressively more low
mass halos up into each richness bin than the fiducial model. To counter this
effect, the mean mass-richness relation drops in amplitude and becomes
shallower, achieving a good match to the original number counts.  However, the
changes in $\ds(R)$  are opposite to those that arise from increasing $\s8$:
because more low mass halos scatter into a given richness bin when $\sc$ is
higher, the amplitude of $\ds(R)$ decreases despite the shift in the mean
richness--mass relation, dropping on both large and small scales. The impact of
increased scatter is higher at larger richness because of the steeper mass
function in this regime.
\end{itemize}
For a more detailed discussion of the dependence of halo populations on $\om$ and
$\s8$, we refer the reader to~\cite{zheng2002}. 

We plot representative measurements and error bars from the S09 $\ds(R)$ data in
the middle panels, specifically the $5\hmpc$ and $15\hmpc$ points in each of the
three richness bins. The impact of the (large) $\om$ and $\s8$ changes
illustrated here is significant compared to these statistical errors, and of
course our full $\ds(R)$ data set has six data points for each of five richness
bins~(see Fig.~\ref{fig:model_demo}), with errors that are only mildly
correlated. The data should thus have substantial constraining power. We can see
that there is degeneracy between $\om$ and $\s8$ as expected, but this
degeneracy is partly broken by the different richness and scale dependence of
the two parameter effects. The impact of increasing the scatter $\sc$ by $50\%$
is much smaller than the impact of similar changes to the cosmological
parameters, and it is strongly richness dependent, essentially vanishing for our
low richness bins. We can therefore anticipate that a rather loose prior on this
nuisance parameter will be enough to avoid degradation of the cosmological
constraints.

The $\ds(R)$ changes at $R\leqslant 2\hmpc$ effectively illustrate the origin
of the R10 cosmological constraints, which use the MaxBCG number counts and the
small scale~(1--halo regime) weak lensing measurements. The different impact of
a $\s8$ change at small and large scales shows why the index $\gamma$ of the
best constrained $\s8\om^\gamma$ combination will be higher for our analysis
than for R10's.

Fig.~\ref{fig:datacontour} presents an overview of this paper's principal
results, showing the 1D posterior distribution for each of the 7 model
parameters~(diagonal panels), and the $95\%$ and $68\%$ confidence regions for
all the parameter pairs~(off--diagonal panels). The prior distributions of
$\beta$ and $\dns$ are also shown on corresponding diagonal panels. We will
refer back to Fig.~\ref{fig:pedagogical} and zoom in on different subsets of
Fig.~\ref{fig:datacontour} multiple times in the following discussion. We are
most interested, of course, on the constraints in the $\om$--$\s8$ plane, but
we must understand their dependence on other parameters.

\subsection{Comparison to WMAP}
\label{sec:vswmap}

\begin{figure*}
\centering
\resizebox{1.00\textwidth}{!}
{\includegraphics{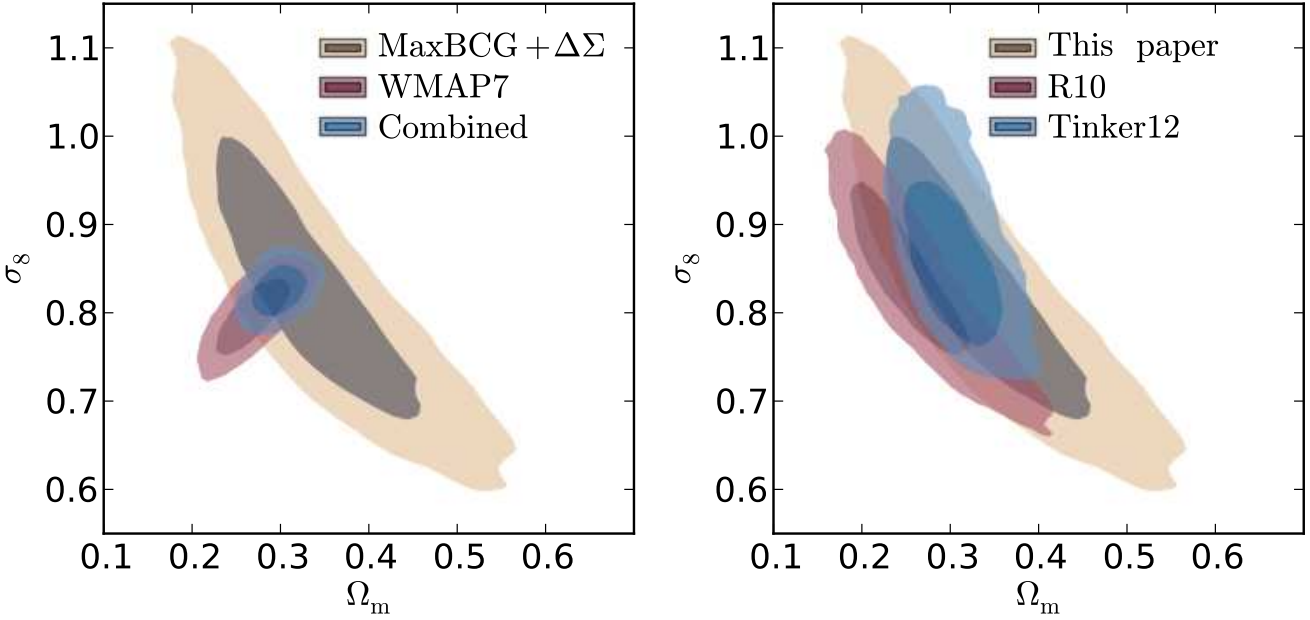}}
\caption{Comparison of our cosmological constraints on the $\om$--$\s8$ plane
with WMAP7 and two other studies using MaxBCG. Left: Constraints from our MaxBCG
analysis~(brown), WMAP7~(red), and the combination of both~(blue). Right:
Constraints from our analysis~(brown), R10~(red), and Tinker12~(blue).}
\label{fig:wmap}
\end{figure*}

The brown contours in the left panel of Fig.~\ref{fig:wmap} show a zoom-in
version of the $\om$--$\s8$ panel from Fig.~\ref{fig:datacontour}, marking the
$68\%$ and $95\%$ confidence limits from our MaxBCG analysis. The error
ellipses are elongated approximately along a degeneracy track of
$\s8(\om/0.325)^{0.501}=0.828\pm0.049$, with the marginalized constraints
$\om=0.325_{-0.067}^{+0.086}$ and $\s8=0.828_{-0.097}^{+0.111}$~($1\sigma$
errors). The $\s8\om^\gamma$ alignment of the error ellipses is typically seen
in cluster abundance--based cosmological constraints, reflecting the
counterbalancing impact of the two parameters on the halo mass function.  The
exact value of $\gamma$ is contingent on cluster mass range and ancillary
information used in each analysis, but it usually lies between $0.4$ and $0.6$.

Our results are consistent with but orthogonal to the {\it WMAP} seven-year
results, which are shown as the red contours in the left panel of
Fig.~\ref{fig:wmap}~\citep{komatsu2011}. Our measurements pull in the direction
of higher $\om$ and $\s8$ relative to WMAP alone. The WMAP constraints rely
strongly on the assumptions of the flat $\lcdm$ cosmological model to
extrapolate growth from $z=1100$ to low redshifts. Our results are only weakly
dependent on $\lcdm$ assumptions, as we are using empirical constraints on the
shape of $\pk$ and measuring a cross--correlation that scales directly with the
matter density and the low redshift amplitude of matter clustering. As with
cluster mass function studies, therefore, our results can be viewed as a
consistency test of the GR $+$ cosmological constant model, one
that focuses on growth of structure rather than expansion history. The joint
constraints from combining the two experiments shrink the regions of equivalent
confidence limits to the blue contours, yielding $\om=0.298_{-0.020}^{+0.019}$
and $\s8=0.831_{-0.020}^{+0.020}$. The best--fit
joint model is summarized in the ``MaxBCG$+$WMAP7'' column of
Table~\ref{tab:bestfit}.

\subsection{Comparison to Other Cluster Cosmology
Probes}
\label{sec:vscluster}

\begin{table}
\centering
\caption{Input data used in the three MaxBCG--based cosmological constraints.}
\begin{tabular}{lccc}
\hline
                     & R10  & Tinker12          & This paper     \\
\hline
Abundance            & Yes  & No                & Yes            \\
Small Scale $\ds(R)$ & Yes  & Yes               & No             \\
Large Scale $\ds(R)$ & No   & No                & Yes            \\ 
Other                & None & Galaxy Clustering & None           \\
\hline
\end{tabular}
\label{tab:r10t12} 
\end{table}

The right panel of Fig.~\ref{fig:wmap} compares the constraints from our
analysis to those of  two other studies using the same cluster sample, R10
and~\citeauthor{tinker2012}~(2012, hereafter Tinker12).  Although using the same
underlying clusters and weak lensing measurements, the three analyses are quite
different from one another, as highlighted in Table~\ref{tab:r10t12}. (Regarding
the fourth row, note that galaxy clustering plays a tangential role in our
analysis via the power spectrum shape prior but is central to the Tinker12
analysis.)

Using the weak lensing masses, R10 obtained constraint
$\s8(\om/0.25)^{0.41}=0.832\pm0.033$~(red contours in the right panel
of~Fig.~\ref{fig:wmap}). The degeneracy index $\gamma=0.41$ is slightly
shallower than the R10 constraint because the small scale $\ds(R)$ responds more
strongly to $\s8$ than the large scale $\ds(R)$, as seen in the central panel of
Fig.~\ref{fig:pedagogical}. The R10 errors are smaller than ours, primarily
because of the higher S/N of the small scale $\ds(R)$ measurements used in their
analysis. It is worth emphasizing, however, that our errors are dominated by
statistical errors in the $\ds(R)$ data, while the R10 errors have substantial
contributions from the weak lensing bias uncertainty $\be$ and from
uncertainties on the mis--centering correlation.

The blue contours in the right panel of~Fig.~\ref{fig:wmap} show the constraint
from Tinker12 expressed as $\s8(\om/0.290)^{0.5}=0.863\pm0.048$. Tinker12
derived constraints from measuring the mass-to-number ratio within MaxBCG
clusters, therefore using the same scales of $\ds(R)$ as R10. They employed
additional information from \citeauthor{zehavi2011}'s (2011) galaxy clustering
measurements, which they used to constrain the halo occupation
distribution~(HOD) within clusters as a function of cosmology. The dominant
uncertainties in the Tinker12 error bars are systematic, from uncertainties in
evolution of the galaxy luminosity function and the galaxy HOD, and from
theoretical uncertainties in the halo mass function and halo bias relation.

In the right panel of Fig.~\ref{fig:wmap}, the three sets of contours largely
overlap with each other, around the region that corresponds to the $68\%$
confidence limit from our WMAP7 joint constraints shown in the left panel. This
good agreement is encouraging, since our analysis uses different scales of
$\ds(R)$ and is insensitive to systematic uncertainties that affect the other
two analyses. There are enough commonalities that it would be risky to combine
the three constraints as though they were independent, but the agreement
certainly suggests that the errors are no larger than those quoted in the
individual studies. As the R10 and Tinker12 results have been
shown to agree with the constraints from X-ray cluster studies, the consistency
between our results and theirs expands the evidence for broad consistency 
between optical and X-ray studies in cluster cosmology. As
statistical precision improves with larger samples~(of clusters, of WL source
galaxies, and of high quality X-ray measurements), these consistency tests
will become considerably more stringent.

\subsection{Constraint on the Richness--Mass Relation}
\label{sec:bestmr}

For specified values of $\om$ and $\s8$, the statistical mass calibration in
our analysis places strong constraints on the richness--mass relation. The
inferred relation shifts systematically with $\om$ and $\s8$, as the
constraint comes mainly from matching the observed richness distribution given
the halo mass function. The model parameters are $\la$ and $\lb$, but the
constraint can be expressed in a more intuitive form by changing parameters to 
\begin{equation}
\langle \ln \rich | \M \rangle = A + \alpha (\ln \M/\M_\mathrm{pivot})
\label{eqn:meanmr}
\end{equation}
where the pivot mass is chosen to minimize correlation between the amplitude
$A$ and slope $\alpha$. Even for fixed cosmology, the values of $A$ and
$\alpha$ are inevitably correlated with the scatter $\sc$, because it is the
combination of the central relation and the scatter that determines the
distribution of richness as a function of mass. As shown in the bottom panel
of Fig.~\ref{fig:pedagogical}, one can reproduce nearly the same cluster number
counts by increasing $\sc$, lowering $A$, and adopting a slightly shallower
slope $\alpha$. However, the richness dependence of $\ds(R)$ provides some
leverage to break this degeneracy.

For the cosmological parameters and scatter of our best--fit MaxBCG$+$WMAP7
model, we obtain $A= 3.233\pm 0.020$ and $\alpha=0.739 \pm 0.009$ with a pivot
mass $\M_\mrm{pivot}=3.614\times10^{14}\hmsun$, which minimizes the correlation
between the two parameters. These ($68\%$) errors are
marginalized over the nuisance parameters $\be$ and $\dns$. If we also
marginalize over $\sc$ but use the same pivot mass, the errors increase to
$A=3.255 \pm 0.069$ and $\alpha=0.744\pm 0.016$. We evaluate the dependence of
$A$ and $\alpha$ on $\om$, $\s8$, and $\sc$ by shifting each parameter in turn
by $\pm 10\%$ from its best--fit MaxBCG$+$WMAP7 value and redetermining the
best--fit $A$ and $\alpha$~(i.e., similar to the experiments described in
Fig.~\ref{fig:pedagogical} but with smaller fractional shifts.). Our final result is 
%om
%amplitude scaling:   -0.15638
%slope     scaling:  -0.011999
%s8
%amplitude scaling:   -0.25363
%slope     scaling:   -0.44637
%sc
%amplitude scaling:  -0.054938
%slope     scaling:   -0.13468
\begin{eqnarray}
A &=& 
\left(\frac{\om}{0.298}\right)^{-0.156}\left(\frac{\s8}{0.831}\right)^{-0.254}\left(\frac{\sc}{0.426}\right)^{-0.055}\nonumber\\
&\times&(3.233 \pm 0.020), \nonumber\\
\alpha &=& 
\left(\frac{\om}{0.298}\right)^{-0.012}\left(\frac{\s8}{0.831}\right)^{-0.446}\left(\frac{\sc}{0.426}\right)^{-0.135}\nonumber\\
&\times&(0.739 \pm 0.009),
\label{eqn:bestrm}
\end{eqnarray}
with pivot mass $\M_\mrm{pivot}=3.614\times10^{14}\hmsun$. Away from the
fiducial values of Equation~\ref{eqn:bestrm}, the errors on $A$ and $\alpha$
may change, and they may become moderately correlated as the effective pivot
mass drifts.

To compare to the best--fit $\langle \ln \rich | \M \rangle$ in R10, we scale
our best--fit $A$ and $\alpha$ to their fiducial cosmology and scatter using the
equations above, and normalize the amplitude to their
$\M_\mathrm{pivot}=7.63\times10^{13}h_{0.7}^{-1}\msun$, the best--fit values are
then $A' = 2.140$ and $\alpha' = 0.890$, consistent with the R10
constraints~($A_\mrm{R10}=2.34\pm{0.10}$ and
$\alpha_\mrm{R10}=0.757\pm{0.066}$). Our richness--mass relation is rotated
upward from the R10 relation by $\simeq 4.5^{\circ}$ around
$3.432\times10^{14}\hmsun$, where the two cross.

We list the average mass and bias of MaxBCG clusters in bins of richness as
computed from the best--fit MaxBCG$+$WMAP7 joint model in
Tables~\ref{tab:abundancebins} and~\ref{tab:dsrbins}. More specifically, we
take the best--fit model parameters and calculate the kernel function weighted
averages via
\begin{eqnarray}
\avg{\M_\mrm{200m}} &=&\frac{1}{\avg{N_{j}}}\int \M \, \wgtj d\M dz \\
\avg{b} &=&\frac{1}{\avg{N_{j}}}\int b(\M, z) \, \wgtj d\M dz.
\end{eqnarray}
Note that the values of $\avg{\M_\mrm{200m}}$ depend on both the scatter and
the slopes of the halo mass function through $\wgtj$, so they are different
than the ``central'' values from directly inverting the mean richness--mass in
Equation.~\ref{eqn:meanmr}. This is most easily seen from the top panel of
Fig.~\ref{fig:weight}, where the actual halos inside each richness bin are
mostly those with masses well below the ``central`` mass implied by the mean
richness--mass relation for that bin. The mean halo masses are less strongly
offset, as shown by the points in that panel, because the more massive clusters
in the bin carry proportionally more weight. Our average cluster masses are in
good agreement with the predictions from R10~(their table 2, column 3), which
in turn are fits to the weak lensing masses measured in \cite{johnston2007} .

\subsection{Constraint on the Nuisance Parameters}
\label{sec:nuisance}

\begin{figure}
\centering
\resizebox{0.47\textwidth}{!}
{\includegraphics{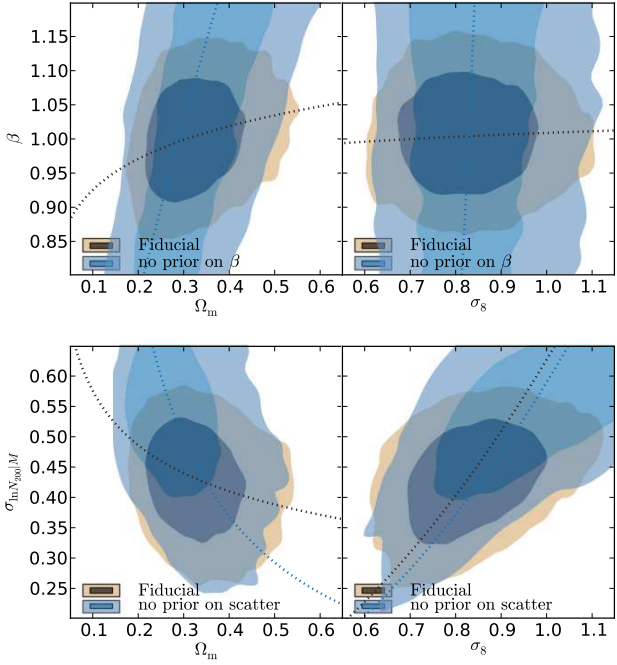}}
\caption{Comparison of the correlations between the cosmological
parameters~($\om$ and $\s8$) and the nuisance parameters~($\sc$ and $\be$),
before and after applying specific priors. In each panel, light and dark brown contours indicate
the $68\%$ and $95\%$ confidence regions from our fiducial analysis, which has
priors on the nuisance parameters, and blue contours show the results when
there is no prior either on $\sc$~(bottom panels) or on $\be$~(top panels). Dotted
curves on top of each contour set represent the degeneracy track followed by the
correlation, calculated from the eigen--decomposition of the
underlying correlation matrices.}
\label{fig:nuisance}
\end{figure}

As identified in Fig.~\ref{fig:pedagogical}, the main nuisance parameters in our
analysis are the scatter in the richness--mass relation $\sc$ and the weak
lensing bias $\be$, both of which are expected to correlate with the
cosmological parameters. To mitigate the impact of these correlations, we
have placed two Gaussian priors to help determine the ranges of $\sc$ and
$\be$, one on the converse scatter and one directly on
$\be$~(Table~\ref{tab:priors}). 

\begin{figure*}
\centering
\resizebox{1.0\textwidth}{!}
{\includegraphics{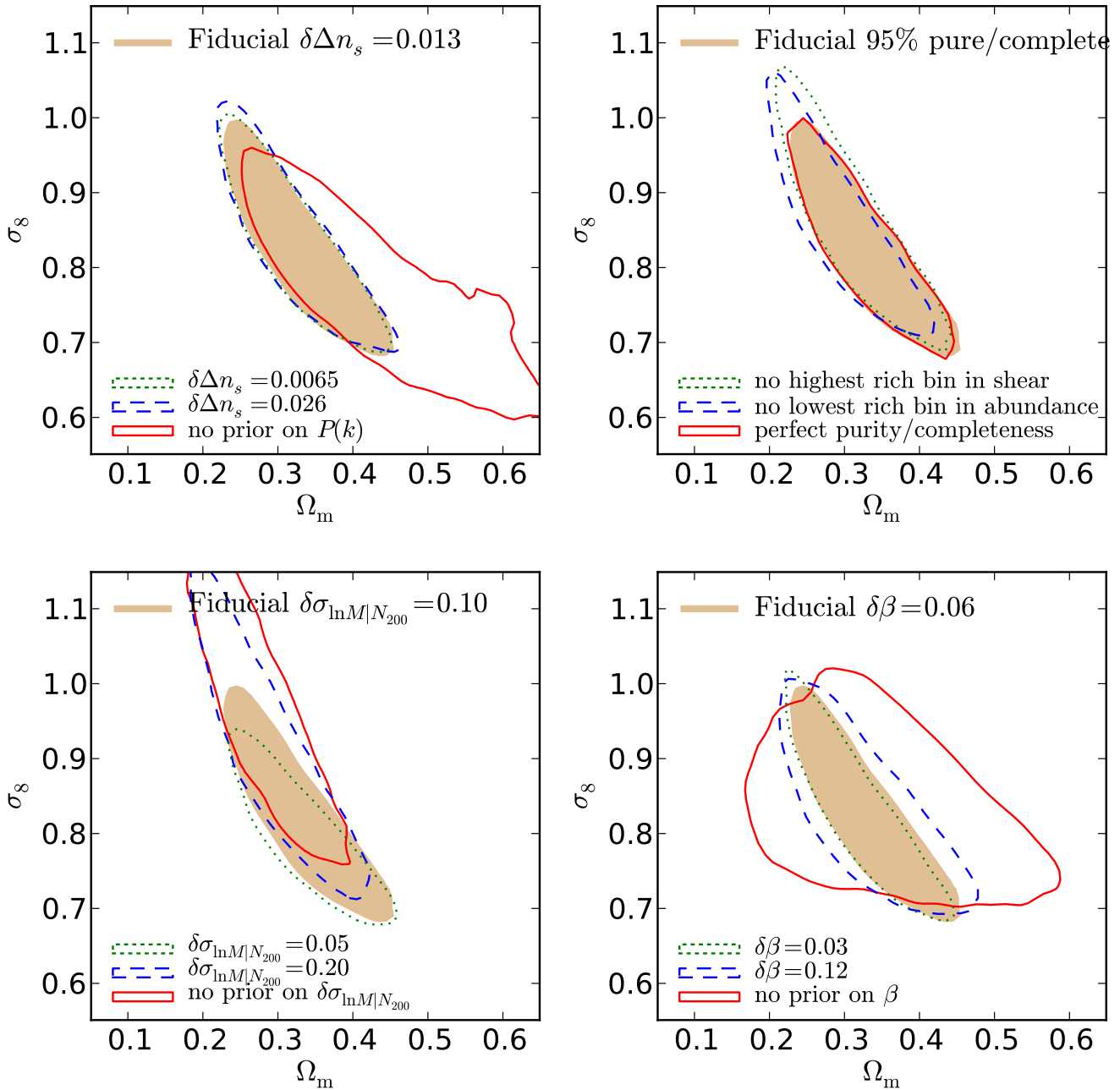}}
\caption{
Effects of relaxing priors and varying assumptions on observational
systematics on the $\om$ and $\s8$ constraint. Contours are $68\%$
confidence regions as constrained by our fiducial analysis~(filled) or different
modifications listed in each panel~(open).}
\label{fig:systematics}
\end{figure*}

\begin{figure*}
\centering
\resizebox{1.0\textwidth}{!}
{\includegraphics{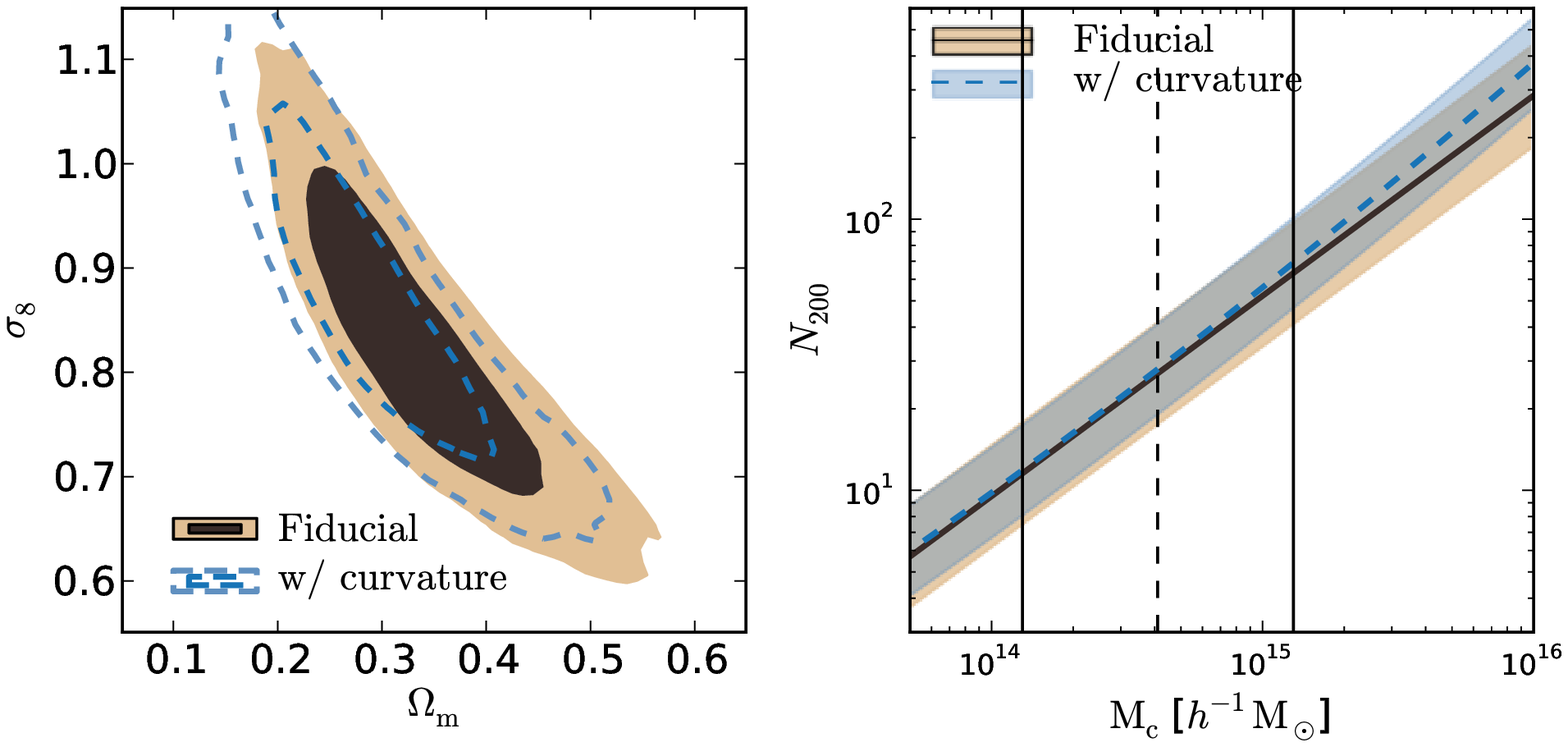}}
\caption{
Effects of allowing curvature in the richness--mass relation on the $\om$
and $\s8$ constraint~(left) and the best--fit curved richness--mass
relation~(right). Left: comparison of the $68\%$ and $95\%$ confidence regions 
derived from the fiducial~(filled) and the curved model~(open). Right:
The best--fit power--law~(thick solid line) and piecewise spline
interpolated~(thick dashed curve) mean richness--mass
relations. The gray band and the thin dashed curves indicate the scatter
about the mean relations. The two
solid vertical lines indicate $\ma$ and $\mb$, while the dashed line indicates
the additional tenor point $\mc$ for the model with curvature .
}
\label{fig:curmr}
\end{figure*}

Fig.~\ref{fig:nuisance} compares the correlations between the two nuisance
parameters ($\sc$ and $\be$) and the two cosmological parameters ($\om$ and
$\s8$), before and after using their respective priors.  As expected, when the prior on
the converse scatter $\scm$ is absent, we observe strong correlation between
$\sc$ and $\s8$, scaling as $\s8(\sc/0.534)^{-0.483}=0.953$~(blue dotted curve in the
bottom right panel). The prior on $\scm$ effectively eliminates the high--$\sc$
and high--$\s8$ region and shifts the residual correlation to
$\s8(\sc/0.432)^{-0.505}=0.828$~(black dotted curve in the bottom right panel).
Similarly, if we allow $\be$ to vary freely, there is correlation between $\om$
and $\be$, scaling as $\be(\om/0.349)^{-0.711}=1.14$~(blue dotted curve in the top left
panel), which then diminishes under the prior on $\be$ to
$\be(\om/0.325)^{-0.069}=1.0$~(black dotted curve in the top left panel). There
are almost no correlations between $\sc$ and $\om$~(bottom left panel) and
between $\be$ and $\s8$~(top right panel), either before or after 
imposing the priors.

The slope of the $\om$--$\be$ correlation in the no prior case, however, is
intriguing~($-0.711$ as shown by the blue dotted curve in the top left panel).
Consider the experiment illustrated in the top panels of Fig.~\ref{fig:pedagogical}, where we
increase  $\om$ while keeping the shape of $\plin$ and the $z=0$ normalization
$\s8$ fixed. At $z=0$, the halo mass function and halo clustering of this
shifted model are nearly identical to those of the original model except for an
overall shift of the mass scale in proportion to $\om$~\citep[see][]{zheng2002}.
Once the richness--mass relation is shifted by this same constant factor, we
expect a nearly identical cluster--mass correlation function at fixed richness.
We therefore expect $\ds(R)\propto\om\xicm$ to shift in proportion to $\om$, so
there should be degeneracy of the form $\be\om^{-1}=\mrm{constant}$, rather
than $\be\om^{-0.711}$. Our clusters have a median redshift of $z_\mrm{med}=0.23$ rather
than zero, and the amplitude $\s8(z_\mrm{med})$ has a small dependence on $\om$
through the linear growth factor. However, the departure from unit slope in
this degeneracy arises primarily because of the dependence of volume element on
$\om$. When $\om$ increases the inferred volume of our cluster survey (defined
by nearly fixed redshift limits and area) decreases, so the linear shift of the
richness--mass relation that keeps the predicted cluster abundance fixed in
comoving $h^3\mpc^{-3}$ in fact predicts a lower number of MaxBCG clusters.

Marginalizing over all other parameters, our constraints on $\sc$ and $\be$ are
$\sc=0.432_{-0.068}^{+0.063}$ and $\be=1.004_{-0.060}^{+0.060}$, respectively,
consistent with the results in R10~($\sc=0.357\pm0.073$ and
$\be=1.016\pm0.060$). Applying WMAP7 priors tightens the constraints to
$\sc=0.436_{-0.024}^{+0.012}$ and $\be=0.968_{-0.030}^{+0.034}$. As we have
already commented in \S\ref{sec:bestmr}, there is mild tension between the
scatter found here and in R10; the difference is only $1\sigma$, but we are
using the same cluster abundance data.

\section{Systematic Errors}
\label{sec:systematics}

We have adopted priors on several of our nuisance parameters, so there could be
systematics beyond our quoted errors if these priors are too tight, or if our
assumption of a power--law richness--mass relation is too restrictive.

Fig.~\ref{fig:systematics} compares the $68\%$ confidence regions in the
$\om$--$\s8$ plane derived from our fiducial analysis~(filled contours) to
those obtained for different priors or variations in the observational
analysis~(open contours).

The bottom left panel explores the robustness of
our analysis against uncertainties in the scatter. When no prior on scatter is
applied, the low--$\om$ and high--$\s8$ regions~(red solid contour) are accepted
because of the degeneracy between $\s8$ and $\sc$ discussed
in~\S~\ref{sec:nuisance}.  Our fiducial prior on the converse scatter, Gaussian
with width $\delta\sc=0.10$, makes an important difference to our individual
errors on $\om$ and $\s8$, though it has little impact on the $\s8\om^{0.501}$
error~(the degeneracy banana gets longer, not wider). If we double the prior
width to $\delta\sc=0.20$~(blue contour), the cosmological constraints are
close to the no prior case. However, if we halve the prior width to
$\delta\sc=0.05$~(green contour),there is only modest tightening of the
constraints; at the current level of statistical error in $\ds(R)$, better
external knowledge of $\sc$ would not make much improvement in our results.
Compared to Fig.~\ref{fig:wmap}, we see that the additional parameter region
allowed by a loose prior on scatter is ruled out if we combine with the
orthogonal WMAP7 constraint. This is the reason that the posterior constraint on
$\sc$ is much tighter when we include WMAP7~(Table~\ref{tab:bestfit}), even
though WMAP does not probe clusters directly.

The bottom right panel of Fig.~\ref{fig:systematics} tests the robustness of
our results against uncertainties in $\be$. As discussed
in~\S~\ref{sec:nuisance}, our internal constraint on $\be$ comes almost
entirely from the volumetric effect of $\om$, so we expect widening/narrowing
of the Gaussian prior on $\be$ to have a much larger impact on $\om$ than on
$\s8$. The blue dashed and green dotted contours show the results after we
double and halve the width of the prior on $\be$, respectively. The blue
contour expands along the $\om$ axis with little change along the $\s8$ axis,
as anticipated. Halving the prior width produces only slight improvement in the
constraints; if our fiducial prior $\delta\be=0.06$ is
accurate-to-conservative, as we think it is, then our constraints are limited
by the statistical errors of the weak lensing measurements rather than the
systematic uncertainties.  If the prior on $\be$ is dropped completely, i.e.,
we allow arbitrary rescaling of the weak lensing measurements and rely only on
the relative $\ds(R)$ amplitudes between bins of different richness, then the
contour expands to fill nearly all possible $\om$ ranges while showing no
degradation of the constraint on $\s8$. 

The top left panel of Fig.~\ref{fig:systematics} addresses the possible
systematics associated with our uncertainties in the $\pks$.  The red solid
contour shows the constraints on $\om$ and $\s8$ without the $\pks$ prior.  As
demonstrated in~\S~\ref{sec:testshape}, this prior helps eliminate the
physically improbable regions of high--$\om$ and low--$\s8$, which is otherwise
favored due to the degeneracy between the $\pks$ and the cosmological
parameters. The blue dashed and green dotted contours represent the $68\%$
confidence regions after increasing and decreasing the width of the Gaussian
prior on $\dns$ by factor of two, respectively. Both contours barely differ
from the fiducial result, indicating our constraints on $\om$--$\s8$ are robust
against the uncertainties in the tilt of the primordial power spectrum within
the range allowed by galaxy surveys. 

In the top right panel of Fig.~\ref{fig:systematics}, we examine the robustness
of our constraint against the uncertainties in the completeness and purity level
of the sample and its sensitivity to the $\ds(R)$ measurements in the extreme
richness clusters. As discussed in \S\ref{sec:covmat}, we allow for potential
biases related to incompleteness or contamination by adding
$\mrm{Var}(\lambda)=0.05^2$ to all elements of our covariance matrix, diagonal
and off--diagonal, based on estimates that the MaxBCG catalog is at least
$95\%$ complete and pure in our richness range. The red contour shows the
effect of dropping this contribution to the covariance matrix, which is
negligible. We conclude that uncertainties in completeness and contamination at
the $5\%$ level do not affect our constraints.

For the richness dependence of $\lambda$, it is known that low richness clusters
are subject to a higher rate of contamination than rich clusters, so we try our
analysis excluding the number count datum for the lowest richness bin at
$\rich\in[11-15]$. The result is shown as the blue dashed contour, which drifts
up from the fiducial one approximately along the degeneracy track. The
constraints from the blue dashed contour are
$\s8(\om/0.293)^{0.489}=0.863\pm0.049$, which slightly torques the halo mass
function to better fit the cluster richness function beyond the lowest bin.
Since the abundance error is smallest for our lowest richness bin, it carries
significant weight in the analysis, so it can have a noticeable impact even
though it is only one of nine abundance data points. However, the drift of the
contour is small compared to its size, so even contamination at the $5\%$ level
in this bin would affect our result at a level small compared to the statistical
error.

As for the stacked $\ds(R)$ measurements, the largest uncertainty occurs at the
highest richness bin where the total number of source galaxies is the least.
The green dotted contour shows the effect of removing the highest richness bin
in the $\ds(R)$ measurements. The resulting confidence region elongates to
accept some high-$\s8$ regions, because despite being noisy, the highest
richness bin carries more $\s8$--sensitive information than other bins.

Although the average mapping between the true mass and optical richness of
clusters should be monotonic, a power--law may be an over--simplification.  To
allow curvature in the mapping, we add a third parameter in the mean
richness--mass relation as the mean log-richness $\lc$ at $\mc =
4.1\times10^{14}\hmsun$~(i.e., the geometric mean of $\ma$ and $\mb$), then we
spline interpolate through the three tenor points on the log-mass vs.
log-richness plane for any given $\la$, $\lb$, and $\lc$, to find a smooth
curve that represents the new mean richness--mass relation.
Fig.~\ref{fig:curmr} shows the results of adding curvature at $\mc$.  Similar
to the test where we dropped the lowest richness bin in
~\S~\ref{sec:systematics}, the confidence regions slide up along the degeneracy
track for a gentle torque in the halo mass function~(blue open contours in the
left panel), and the richness--mass relation bends slightly~(blue dashed curve
in the right panel) to reduce the number of the highest richness clusters.  In
this way, the model achieves a better fit to the detailed shape of the observed
richness function than the fiducial power--law model. However, the best--fit
curved richness--mass relation remains very close to a power law over the
relevant mass range, and the parameter constraints shift only slightly relative
to the fiducial ones. The width of the constraints grows only small amount,
indicating that our assumption of a power--law relation in our fiducial
analysis does not bias or overly restrict our results.

\section{Summary and Future Prospects}
\label{sec:dis}

We have derived cosmological constraints on $\om$ and $\s8$ using the
combination of large scale cluster--galaxy weak lensing measurements~(S09) and
the abundance of MaxBCG clusters as a function of richness. Within the
analysis, we have statistically calibrated the cluster masses by requiring
consistency between the cosmological model fit and the data, exploiting
external priors on the scatter in the richness--mass relation from comparisons
to X-ray data~\citep{rozo2009}, and on the
$\pks$ from galaxy clustering~\cite{reid2010}. The $68\%$ confidence ellipse of our cosmological
constraints on the $\om$--$\s8$ plane can be summarized as 
\begin{equation}
\s8(\om/0.325)^{0.501}=0.828\pm0.049, 
\end{equation}
which is consistent with and orthogonal to the WMAP7 constraints on these
parameters. This consistency of structure measured in the recombination era and
the low redshift universe provides further evidence for the gravitational
growth predicted by the $\lcdm$ model combining GR, a cosmological constant,
and cold dark matter. Assuming this model to be correct and combining our
analysis with WMAP7, we obtain individual constraints as
\begin{equation}
\om = 0.298\pm{0.020} \quad\mbox{and}\quad
\s8 = 0.831\pm{0.020} .
\end{equation}
The overall results are consistent with and complementary to two other
cosmological constraints from the same underlying clusters but with different
input data and systematic uncertainties~(R10 and Tinker12). Collectively
these three studies demonstrate consistency of the small scale weak lensing,
large scale weak lensing, galaxy content, and abundance of the MaxBCG sample,
together with galaxy clustering data.

The primary systematic uncertainties in our analysis are the scatter in the
cluster richness--mass relation and residual bias in the weak lensing
measurements associated with photometric redshifts or shear calibration.
However, with the external priors we have adopted, neither of these systematics
is a limiting factor in our analysis; the uncertainties in our cosmological
constraints are dominated by statistical uncertainties in the large scale
$\ds(R)$ measurements. These statistical errors can be sharply reduced in
future surveys with deeper imaging and better seeing. Statistical improvements
will require corresponding improvements in the control of systematics. While we
have focused in this paper on large scales to complement the R10 analysis, the
long term goal should be to derive constraints from the full range of $\ds(R)$
simultaneously. Achieving this goal will require theoretical and numerical work
to construct models that are accurate across the 1-halo and 2-halo transition
and to assess uncertainties in the accuracy of the model predictions at all
scales.

There are opportunities for significant near--term improvements in our analysis
using SDSS data. The MaxBCG catalog and weak lensing measurements used here are
based on imaging data from DR4. With the recent release of
DR8~\citep{aihara2011}, almost every aspect of the catalog construction and the
weak lensing measurements has evolved. 
The increase in the imaging area will enhance the raw statistical power~(for
 $7,398$ deg$^2$ VS. $14,555$ deg$^2$),
reducing Poisson uncertainties and sample variance in the cluster counts and
shape noise in the $\ds(R)$ measurements. The optical cluster finding algorithm
has been improved to produce catalogs with well-controlled selection function
and, more importantly, a new richness estimator with reduced intrinsic scatter.
\cite{rykoff2012} considered various modifications of the original richness
estimator in MaxBCG and found that the scatter in log--mass at fixed richness
could be reduce to $0.2-0.3$ depending on richness, substantially smaller than
MaxBCG scatter~\citep[$0.45\pm0.10$;][]{rozo2009} that we adopted as a prior in
our analysis. When the scatter itself is smaller, then the systematic
uncertainty tied to uncertainty in the scatter is also smaller. With improved
uncertainty of the selection function, it will be feasible to use higher
redshift clusters in the analysis, and while the $\ds(R)$ measurements will
degrade at higher $z$ because of reduced source surface density, the leverage
of a wider redshift range may strengthen the cosmological constraints.  On the
weak lensing side, the main improvement is a better understanding of the
photometric redshift distribution of source galaxies. With a much improved
spectroscopic training set and better photometric calibration,
\cite{sheldon2011} reconstructed a redshift distribution for DR8 imaging data
that is primarily limited by sample variance.  Additional improvements come
from updates in the photometric pipeline, including better sky subtraction,
more refined stellar masks, and better PSF corrections in the shape
measurements.

Beyond SDSS, our approach can be applied to future, deeper, large--area imaging
surveys. In the near term, the Pan--STARRS1~\cite[PS1;][]{chambers2007} $3\pi$
survey is expected to have larger area than SDSS, slightly greater depth, and
higher image quality that yields a significant increase of the source density
for weak lensing. The Dark Energy
Survey~\cite[DES;][]{the_dark_energy_survey_collaboration2005}, expected to
start in late 2012, plans to survey $5000$ deg$^2$ to a depth two magnitudes
beyond SDSS, with a weak lensing source density a factor of ten higher. It is
designed with cluster cosmology and weak lensing as central goals, and our
technique is naturally adapted to it. In the longer term, the imaging data sets
from LSST, and the Euclid and WFIRST missions will allow radical improvements in
the precision of cluster--galaxy lensing analysis, with effective source
densities of $20-40$ arcmin$^{-2}$. These imaging surveys can provide their own
cluster catalogs identified from the galaxy population, and they can provide
stacked weak lensing measurements for clusters identified via X-ray emission or
the SZ effect. Comparisons of results from different classes of cluster catalogs
allow powerful cross--checks for systematics~\citep[see, e.g.,][]{rozo2012,
rozo2012-1, rozo2012-2} and valuable constraints on mass--observable
relations~\citep{cunha2010}.  The SZ effect is a powerful technique for finding
massive clusters at very high redshifts~\citep[e.g.,][]{reichardt2012,
williamson2011, marriage2011}, and DES will target the area covered by the SZ
survey of the South Pole Telescope. X-ray observables may be more tightly
correlated with halo mass than optical observables, and the upcoming eROSITA
mission will carry out a sensitive all--sky X-ray survey that will revolutionize
cosmological studies with X-ray selected clusters.

\cite{oguri2011} and \cite{weinberg2012} argue that cluster abundances with
masses calibrated by stacked weak lensing can provide constraints on structure
growth that are highly competitive with those from cosmic shear analysis of the
same WL survey.  This conclusion assumes $\ds(R)$ measurements out to $\sim 1-2$
cluster virial radii, so the larger scale analysis illustrated here can only
strengthen the power of this approach.  Cluster-galaxy lensing is analogous to
galaxy-galaxy lensing, but the relation between clusters and halos is simpler
than the relation between galaxies and halos, and it is less subject to
the complexities of baryonic physics.  This greater simplicity reduces systematic
uncertainty associated with theoretical modeling, which may ultimately
compensate for the rarity of clusters relative to galaxies (and consequent lower
statistical precision of the WL measurements).  The Tinker12 study shows that
the constraining power of small scale measurements can be enhanced by bringing
in additional information from galaxy clustering and cluster mass-to-number
ratios.  In future work, we will investigate the generalization of this idea to
large scales using the cluster-galaxy cross-correlation function, which can be
measured in projection from the same survey used for weak lensing analysis.
Nature has provided observable signposts that mark the locations of the most
massive halos in the universe, and stacked weak lensing provides a tool to
measure the average mass profiles of these halos at high precision over a wide
range of scales.  Exploiting this combination promises to yield stringent tests
of gravity on cosmological scales and of theories for the origin of cosmic
acceleration.

\section*{Acknowledgements}

We thank Anatoly Klypin for providing the L1000W simulation. D.H.W. and Y.Z are
supported by the NSF grant AST-1009505. E.R. is funded by NASA through the
Einstein Fellowship Program, grant PF9-00068. M.R.B. is supported in part by the
Kavli Institute for Cosmological Physics at the University of Chicago through
grants NSF PHY-0114422 and NSF PHY-0551142, and an endowment from the Kavli
Foundation and its founder Fred Kavli.

Funding for the SDSS and SDSS-II has been provided by the Alfred P. Sloan
Foundation, the Participating Institutions, the National Science Foundation,
the U.S. Department of Energy, the National Aeronautics and Space
Administration, the Japanese Monbukagakusho, the Max Planck Society, and the
Higher Education Funding Council for England. The SDSS Web Site is
http://www.sdss.org/.

The SDSS is managed by the Astrophysical Research Consortium for the
Participating Institutions. The Participating Institutions are the American
Museum of Natural History, Astrophysical Institute Potsdam, University of
Basel, University of Cambridge, Case Western Reserve University, University of
Chicago, Drexel University, Fermilab, the Institute for Advanced Study, the
Japan Participation Group, Johns Hopkins University, the Joint Institute for
Nuclear Astrophysics, the Kavli Institute for Particle Astrophysics and
Cosmology, the Korean Scientist Group, the Chinese Academy of Sciences
(LAMOST), Los Alamos National Laboratory, the Max-Planck-Institute for
Astronomy (MPIA), the Max-Planck-Institute for Astrophysics (MPA), New Mexico
State University, Ohio State University, University of Pittsburgh, University
of Portsmouth, Princeton University, the United States Naval Observatory, and
the University of Washington.

%%%%%%%%%%%%%%%%%%%%%%%%%%%%%%%%%%%%%%%%%%%%%%%%%%%%%%%
%  Bibliography
%%%%%%%%%%%%%%%%%%%%%%%%%%%%%%%%%%%%%%%%%%%%%%%%%%%%%%%
%\clearpage

%\bibliographystyle{apj}
%\footnotesize{
%\bibliographystyle{mn2e}
%\bibliography{clusterwl-paper} 
%}

\footnotesize{

}

%%%%%%%%%%%%%%%%%%%%%%%%%%%%%%%%%%%%%%%%%%%%%%%%%%%%%%%
%
%%%%%%%%%%%%%%%%%%%%%%%%%%%%%%%%%%%%%%%%%%%%%%%%%%%%%%%
\end{document}